\documentclass[%
 pre,
 amsmath,amssymb,
 reprint,%
]{revtex4-1}

\usepackage{amsmath}
\usepackage{amssymb}
\usepackage{float}
\usepackage[utf8]{inputenc}
\usepackage{hyperref}

\usepackage{tikz}
\usetikzlibrary{calc,decorations.markings}
\usepackage{subcaption}

\usepackage{colortbl}
\usepackage{lipsum}

\DeclareMathAlphabet\mathbfcal{OMS}{cmsy}{b}{n}

\begin{document}

%\preprint{AIP/123-QED}

\title{Angular momentum gain by electrons under action of intense structured light}
%\author{N. Bukharskii}
%\altaffiliation[Also at ]{P.N. Lebedev Physical Institute of RAS, 53 Leninskii Prospekt, 119991 Moscow, Russian Federation.}
\author{E. Dmitriev}
% \altaffiliation[Also at ]{P.N. Lebedev Physical Institute of RAS, 53 Leninskii Prospekt, 119991 Moscow, Russian Federation.}
\author{Ph. Korneev}
 \email{korneev@theor.mephi.ru; ph.korneev@gmail.com}
% \altaffiliation[Also at ]{P.N. Lebedev Physical Institute of RAS, 53 Leninskii Prospekt, 119991 Moscow, Russian Federation.}
\affiliation{ 
%National Research Nuclear University MEPhI, 31 Kashirskoe shosse, 115409 Moscow, Russian Federation
P.N. Lebedev Physical Institute of RAS, 53 Leninskii Prospekt, 119991 Moscow, Russian Federation.
}

\date{\today}% It is always \today, today,

\begin{abstract}
The problem of light waves interaction with charged particles becomes more and more complex starting with the case of plane waves, where the analytical solution is well known, to more natural, though more complicated situations which include focused or structured laser beams. Internal structure may introduce a new degree of freedom and qualitatively change the dynamics of interacting particles. For certain conditions, namely for the dilute plasma, description of single-particle dynamics in the focused structured laser beams is the first step and may serve as a good approximation on the way of understanding the global plasma response. Moreover, the general problem of integrability in complex systems starts from consideration of the integrals of motion for a single particle. The primary goal of this work is an understanding of the physics of the orbital angular momentum (OAM) absorption by a single particle in a focused structured light. A theoretical model of the process, including solutions of Maxwell equations with the required accuracy and a high-order perturbative approach to electron motion in external electromagnetic fields, is developed and its predictions are examined with numerical simulations for several exemplary electromagnetic field configurations. In particular, it was found that for the particles distributed initially with the azimuthal symmetry around the beam propagation direction, the transferred OAM has a smallness of the fourth order of the applied field amplitude, and requires an accurate consideration of the temporal laser pulse envelope.    
\end{abstract}

\maketitle

  \section{Introduction}

%\bb{Egor, need to make a brief literature survey. Collect about 10-20 main papers on the OAM light interaction with charged particles and write 2-3 sentences describing what is done in each one.}

Electromagnetic interaction between structured light and charged particles in vacuum may result in particle acceleration, i.e. momentum and energy transfer, which under certain conditions appears to be irreversible. For that, the light wave should at least be sufficiently different from a slow-varying plane wave, see, e.g. \cite{LL2-eng}. In this general context, a similar question may be posed concerning the transfer of the light orbital angular momentum (OAM): how effective it occurs when a structured light wave interacts in vacuum with a single charged particle, and how effective it is in average for a particle ensemble with a given distribution.

This question was addressed recently within different frameworks. In \cite{tikhonchuk2020numerical, toma2023scaling, molnar2021direct}, the OAM transfer is considered numerically for different model configurations of laser beams interacting with individual charged particles in free space. Paraxial and slowly varying envelope approximations were used to prescribe the electromagnetic fields, and specific cases of linear, circular and radial polarizations with and without spatio-temporal coupling were considered. Transferred energy, momentum and angular momentum dependence on duration, amplitude and other parameters were studied in some certain cases. In Ref. \cite{dmitriev2023laser}, numerical simulation of the action of the OAM laser beam on free electrons is presented for the case of superposition of a linearly polarized Laguerre-Gaussian and Gaussian modes, the OAM transfer is estimated based on the results of the perturbation theory. An importance of the accurate use of the paraxial and slowly varying temporal envelope approximations in numerical simulations is discussed in \cite{dmitriev2022effect} based on consideration of a conserved integral of motion for the special case, when the circular polarization and orbital momentum in the beam have opposite directions, so that there is no dependence on the angle in the phase. The necessity of corrections to the lowest order approximations both in numerical simulations and analytics is demonstrated.  

More complicated, though less detailed studies of the OAM beam interaction with plasmas, were mainly performed with the use of large-scale 3D particle-and-cell simulations. In Refs. \cite{nuter2018plasma, nuter2020gain}, generation of magnetic fields in dilute plasma was observed and explained as a result of an OAM transfer from the laser beam to electrons. There, radial and linear polarizations were considered, showing effective OAM transfer for moderate relativistic intensities. Analysis with the use of simplified electromagnetic fields showed the importance of the longitudinal particle motion for the OAM transfer. Magnetic field was also observed in plasma in Ref. \cite{shi2018magnetic} via OAM transfer from two beating Laguerre-Gaussian beams to charged particles. There, a fluid model of the interaction was developed to describe formation of azimuthal currents in the plasma. 

In case of dissipative processes, the magnetic field generation, originating from angular momentum absorption of a circularly polarized laser beam, is demonstrated theoretically in \cite{berezhiani1997theory}. Circularly polarized beams were replaced by linearly polarized beams with OAM in Ref. \cite{ali2010inverse}. For intense relativistic beams with different polarizations, including linearly and circularly polarized Laguerre-Gaussian modes, Ref. \cite{longman2021kilo} studies magnetic field generation and dissipative effects in plasma. 
%OAM transfer process in a plasma channel formed by a Laguerre-Gaussian beam is discussed in \cite{punia2020generation}. Analytical model, based on resonant electron acceleration is developed.

Although there is a common conclusion that the OAM transfer from a structured light wave to electrons may be both possible and effective, still, it is not clear if this is always a single-particle effect or certain nonlinearities originated from collective or dissipative effects are required. It was observed in several studies \cite{nuter2018plasma, toma2023scaling}, that the individual electrons are absorbing positive or negative light OAM depending on their spatial position, even for definite OAM sign in the incident wave, but the net gain appears to be much less. The detailed consideration for the analytically attainable special case, predicting no net OAM gain after interaction \cite{dmitriev2022effect}, supported the conclusion that the total OAM transfer is a very delicate process.   

In this work, a general problem of OAM transfer from a structured light wave to an ensemble of charged particles is considered for moderate intensities of the incident light. It may be a rarefied plasma, consisting of ions and electrons, which are interacting with an incident  light much stronger than between themselves, which requires $\omega_p~\ll~\omega_0$, where $\omega_p=\sqrt{\frac{4\pi n_e e^2}{m_e}}$ is the plasma frequency, $n_e$ is the electron density, $e$ and $m_e$ are the electron charge and mass respectively, $\omega_0$ is the characteristic light frequency, e.g. the main carrier frequency of the laser wave. The particles are considered initially cold with the temperature $T \ll (eE_0)^2/m_e\omega_0$, where $E_0$ is the electric field amplitude. This condition means that the work, performed by the laser wave during one period is small compared to the thermal energy of the electrons, and allows to treat electrons as being at rest before the interaction. Electrons, being the lighter particles, are primarily affected by the light, so the OAM transfer would be analyzed having in mind namely the wave-electron interaction, but of course all the results are valid for any charged particles.

First, a perturbation theory on the field strength for arbitrary fields is developed up to the fourth order, to obtain a nonzero net absorbed OAM in case of homogeneous distribution of the particles. Then, assuming a focused laser beam, an approximate description of the structured wave within the paraxial and the slow temporal dependence approximations is represented. Using this description, several certain configurations are considered in details, including comparison with numerical single-particle calculations. Closer to the end, general discussion and conclusions are presented.

\section{Analytical model}

%  \section{Initial equations}
  
%  \bb{In the main text, leave all the parameters ($c$, $m_e$, $e$); in the appendixes may use $c=1$, $m_e=1$, $e=1$ but then reconstruct in the final expression. Or use parameters also in the appendicies.}
  
  Electron motion in arbitrary electromagnetic fields $\mathbf E$ and $\mathbf H$ is described by the equation
  \begin{equation} \label{dim_eqs_motion}
    \frac{d\mathbf p}{dt} = - e \left( \mathbf E + \frac{\mathbf v}{c} \times \mathbf H \right)
  \end{equation}
  with initial conditions
  \begin{equation}\label{initial_conditions}
    \left\{
    \begin{aligned}
        \mathbf r \left(t \to -\infty\right) & = \mathbf r_0, \\
        \mathbf v \left(t \to -\infty\right) & = 0,
    \end{aligned}
    \right.
  \end{equation}
  where $\mathbf p = m_e \gamma \mathbf v$ is the electron momentum, $\gamma =1/{\sqrt{1 - \frac{v^2}{c^2}}}$ is the Lorentz factor, %$e$ is the electron charge, $c$ is the speed of light, $m_e$ is the electron mass, 
  $\mathbf r_0$ is the electron initial position, $c$ is the speed of light and $\mathbf E = \mathbf E (\mathbf{r},t)$ and $\mathbf H = \mathbf H (\mathbf{r},t)$ are the laser electric and magnetic fields respectively, satisfying Maxwell equations in vacuum
  \begin{equation} \label{Maxwell equations}
  \begin{cases}
      \nabla \times \mathbf E = - {\displaystyle\frac{1}{c}} \displaystyle\frac{\partial \mathbf H}{\partial t}, & \nabla \times \mathbf H = \displaystyle\frac{1}{c} \displaystyle\frac{\partial \mathbf E}{\partial t}, \\
      \nabla \cdot \mathbf H = 0, & \nabla \cdot \mathbf E = 0.
    \end{cases}
  \end{equation}
  \begin{figure}[H] 
      \centering 
      \includegraphics[width = \linewidth]{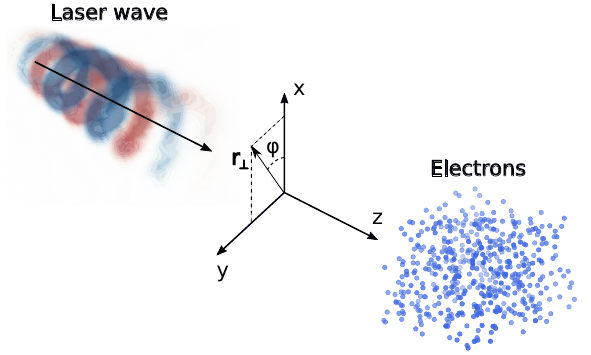} 
      \caption{The interaction scheme and the coordinate system. The $z$ axis coincides with the laser propagation axis and particles are distributed isotropically in the transverse plane $xy$ for every value of $z$.}
      \label{setup}
  \end{figure}
  The schematic setup of the interaction is presented in Fig. \ref{setup}.
   In the following part of the paper a perturbation theory of angular momentum transfer from electromagnetic wave to charged particles is developed.

\subsection{Particle motion in electromagnetic wave}
  
  Consider formally a low intensity regime $a_0 \lesssim 1$, where $a_0 = \frac{e E_0}{m_e\omega_0 c}$ is the dimensionless amplitude of the field, $E_0$ is the amplitude of electric field.
  %, $\omega_0$ is the main carrier frequency.
  The electromagnetic wave is assumed to be finite, $\mathbf E (\mathbf r, t \to \pm \infty) = \mathbf H (\mathbf r, t \to \pm \infty) = 0$. In frames of the perturbation theory on $a_0$, the coordinates and velocities of the particle may be expressed in the form
  \begin{equation}
    \left\{
    \begin{aligned}
        \mathbf r & = \mathbf r^{\left(0\right)} +  \mathbf r^{\left(1\right)} + \mathbf r^{\left(2\right)} + \dots, \\
        \mathbf v & = \mathbf v^{\left(0\right)} +  \mathbf v^{\left(1\right)} + \mathbf v^{\left(2\right)} + \dots,
    \end{aligned}
    \right.
  \end{equation}
  where $\mathbf r^{\left(0\right)}$ and $\mathbf v^{\left(0\right)}$ are the unperturbed coordinate and velocity, $\mathbf r^{\left(n\right)}\sim a_0^n$ and $\mathbf v^{\left(n\right)}\sim a_0^n$. According to the initial conditions \eqref{initial_conditions} $\mathbf r^{\left(0\right)} = \mathbf r_0$, $\mathbf v^{\left(0\right)} = 0$.

  In the first order
  \begin{equation}
    \frac{d\mathbf p^{\left(1\right)}}{dt} = - \frac{m_e c \omega_0}{E_0} a_0 \mathbf E\left(\mathbf r_0, t \right).
  \end{equation}
  %where $\mathbf E \equiv \mathbf E \left(\mathbf r_0, t \right)$ is the electric field, taken at the initial position of the particle.
  The solution of the equation of the first order is
  \begin{equation}
    \left\{
    \begin{aligned}
      \mathbf p^{\left(1\right)} & = m_e \mathbf v^{\left(1\right)} = - \frac{m_e c \omega_0}{E_0} a_0 \int \limits_{-\infty}^{t}d t^\prime \mathbf E\left(\mathbf r_0, t^\prime \right), \\
      \mathbf r^{\left(1\right)} & = - \frac{c \omega_0}{E_0} a_0 \int \limits_{-\infty}^{t} d t^\prime \int \limits_{-\infty}^{t^\prime} d t^{\prime \prime}\mathbf E\left(\mathbf r_0, t^{\prime \prime} \right).
    \end{aligned}
    \right.
  \end{equation}
  The gained momentum of a particle in the first order is
  \begin{equation} \label{p1_inf}
    \mathbf p^{\left(1\right)}\left({t_\infty}\right) = - \frac{m_e c \omega_0}{E_0} a_0 \mathbf E \left( \omega = 0, \mathbf r_0\right) = 0,
  \end{equation}
  where the notation $"t_\infty"\equiv "t\to\infty"$ is used and $\mathbf E(\omega, \mathbf r)$ is the Fourier component of the electric field.
  It is assumed that the electromagnetic pulse has no constant part and hence a particle does not gain momentum in the first order.
  %\bb{the following is not true e.g. for a plane wave, if it is not needed for something else, just omit this statement.}
  Assuming the same for double integration over time, obtain 
  \begin{equation} \label{condition r1}
    \mathbf r^{\left(1\right)}\left({t_\infty}\right) = 0,
  \end{equation}
the validity of this assumption is discussed in more details in Section \ref{discussion}.
  
  According to $\eqref{p1_inf}$, angular momentum in the first order on $a_0$ is not gained by a particle,
  \begin{equation}
      \mathbf L^{\left(1\right)} = \\
      \mathbf r_0 \times \mathbf p^{\left(1\right)} \xrightarrow[{t_\infty}]{} 0.
  \end{equation}
  In the second order perturbation theory
  \begin{multline} \label{dp2dt}
    \frac{d\mathbf p^{\left(2\right)}}{dt} = m_e \frac{d\mathbf v^{\left(2\right)}}{dt} = 
     - \frac{m_e c \omega_0}{E_0} a_0 \times \\ \times \left(\left(\mathbf r^{\left( 1 \right)} \cdot \nabla_0\right)\mathbf E \left(\mathbf r_0, t \right) + \frac{ \mathbf v^{\left( 1 \right)}}{c} \times \mathbf H \left(\mathbf r_0, t \right) \right), 
  \end{multline}
  where $\nabla_0 = \mathbf e_x \frac{\partial}{\partial x_0} + \mathbf e_y \frac{\partial}{\partial y_0} + \mathbf e_z \frac{\partial}{\partial z_0}$ is the del operator, taken with respect to the initial position of the particle $\mathbf r_0$.

 To obtain general expressions for arbitrary electromagnetic fields, consistent with the Maxwell equations, the perturbation theory uses the relations between the electric and magnetic field components. For $\nabla_0 \times \mathbf v^{\left(1\right)}$, taking into account that $\nabla \times \mathbf E = - \frac{1}{c}\frac{\partial \mathbf H}{\partial t}$ one obtains
  \begin{multline}
      \nabla_0 \times \mathbf v^{\left(1\right)} = - \frac{c \omega_0}{E_0} a_0 \int \limits_{-\infty}^{t} \nabla_0 \times \mathbf E\left(\mathbf r_0, t^\prime \right) dt^\prime = \\
     = \frac{\omega_0}{E_0} a_0\int \limits_{-\infty}^{t} \frac{\partial \mathbf H\left(\mathbf r_0, t^\prime \right)}{\partial t^\prime} dt^\prime =  \frac{\omega_0}{E_0} a_0\mathbf H\left(\mathbf r_0, t \right),
  \end{multline}
and substituting $\mathbf H = \frac{E_0}{\omega_0 a_0} \nabla_0 \times \mathbf v^{(1)}$ into \eqref{dp2dt}
  \begin{equation}
    \frac{d}{dt} \left( \mathbf p^{\left(2\right)} - m_e \left(\mathbf r^{\left( 1 \right)} \cdot \nabla_0 \right)\mathbf v^{\left( 1 \right)}\right) = - \frac{m_e}{2} \nabla_0  \left(\mathbf v^{\left(1\right)}\right)^2.
  \end{equation}
The gained momentum in the second order after interaction then reads
  \begin{equation}
    \mathbf p^{\left( 2 \right)} \left( {t_\infty}\right) = -\frac{m_e}{2} \nabla_0 \int \limits_{-\infty}^{\infty} \left(\mathbf v^{\left( 1 \right)} \right)^2 dt,
  \end{equation}
  where it is assumed that $\frac{\partial}{\partial x_i}\mathbf v^{(1)}({t_\infty}) \to 0$.
%  An infinite isotropic system as a whole does not gain momentum in the second order
%  \begin{equation}
%    \int \mathbf p^{\left( 2 \right)} \left( {t_\infty}\right) dV = -\frac{1}{2} \int \nabla\left( \int \limits_{-\infty}^{\infty} \left(\mathbf v^{\left( 1 \right)}\right)^2 dt\right)dV = -\frac{1}{2} \oint \left( \int \limits_{-\infty}^{\infty} \left(\mathbf v^{\left( 1 \right)}\right)^2 dt\right) d \mathbf S = 0,
%  \end{equation}
  The particle displacement in the second order perturbation theory then reads
  \begin{multline} \label{r2}
    \mathbf r^{\left(2\right)} =  \int \limits_{-\infty}^t dt \left(\mathbf r^{\left( 1 \right)} \cdot \nabla_0\right)\mathbf v^{\left( 1 \right)} - \\ -\frac{1}{2} \int \limits_{-\infty}^t dt^\prime \int \limits_{-\infty}^{t^\prime}dt^{\prime \prime} \nabla_0 \left(\mathbf v^{\left(1\right)}\right)^2.
  \end{multline}

  The second-order absorbed angular momentum is
  \begin{equation}
    \mathbf L^{\left(2\right)} =  \mathbf r_0 \times \mathbf p^{\left(2\right)} + \mathbf r^{\left(1\right)} \times \mathbf p^{\left(1\right)}.
  \end{equation}
  Introduce cylindrical coordinates $(r_\perp, \varphi, z)$ with $r_\perp$ and $\varphi$ being the transverse distance and the azimuthal angle respectively, i.e. $x = r_\perp \cos\varphi$, $y = r_\perp \cos\varphi$. Axis $z$ is chosen in such a way that it coincides with the axis of propagation of the electromagnetic wave, see Fig. \ref{setup} and Section \ref{EM_fields} for further details related to the fields. The longitudinal component of the angular momentum after interaction reads
  %may be written in the form \bb{where are the other terms? For arbitrary fields they all should be here, for $t<\infty$. And then go to the limit $t\to\infty$, using \eqref{p1_inf}. }
  \begin{equation}
    L_z^{(2)} ({t_\infty}) =  r_0 p^{\left(2\right)}_\varphi ({t_\infty}) = -\frac{m_e}{2} \frac{\partial }{\partial \varphi_0} \int \limits_{-\infty}^{\infty} \left(\mathbf v^{\left( 1 \right)}\right)^2 dt,
  \end{equation}
where $\mathbf r_0\equiv(r_{\perp 0},\varphi_0,z_0)\equiv(r_0,\varphi_0,z_0)$.
%\bb{define cylindrical system somewhere: where $\mathbf r_0 = (r_0,\varphi_0,z_0)$ in cylindrical coordinates. It is also a question how to define $z$-direction: for arbitrary fields as here it is not clear, for paraxial approximation should be linked to the corresponding equation. Somewhere in the main part we should explain the geometry, define axes and show a scheme.}
%  \textcolor{red}{explain zero subscript}
  After averaging over the azimuthal angle
  \begin{equation}\label{Lz2}
    \langle L_z^{(2)} ({t_\infty})\rangle_{\varphi_0} \equiv \int \limits_0^{2 \pi} \rho(r_0, z_0) L_z^{(2)} ({t_\infty}) d \varphi_0 = 0,
  \end{equation}
  where the distribution function of the initial coordinates of the particles $\rho(r_0, z_0)$ is considered to be symmetric about the laser propagation axis, i.e. it does not depend on the initial angle of the particle $\varphi_0$.
  For instance, the distribution function for an isotropic plasma cylinder is $\rho(r_0, z_0) = 1 / (\pi R^2 h)$ for $r_0, z_0$ that are inside the cylinder and $0$ otherwise, where $R$ is the radius and $h$ is the height of the cylinder respectively.

  As it follows from \eqref{Lz2}, for the plasma, which is isotropic transversely to the wave propagation direction, there is no net angular momentum gain up to the second order of the perturbation theory on $a_0$;
  %\bb{For which fields? The fields are required to define $z$-axis, and for these fields the net $L_z\to 0$. But with some special definition of the coordinates and the fields in that coordinates, a simple momentum gain means angular momentum gain, consider a beam along $x$ which is not symmetrical on $\phi$.}
  to describe the angular momentum transfer higher orders of perturbation theory are required.
  
\begin{widetext}
  The third order perturbation theory gives for the momentum
  \begin{multline}	
    \frac{d\mathbf p^{\left(3\right)}}{dt} = - \frac{m_e c \omega_0}{E_0} a_0 \times \\ \times\left(\left(\mathbf r^{\left( 2 \right)} \cdot \nabla_0\right)\mathbf E\left(\mathbf r_0, t \right) + \frac{1}{2} r^{\left( 1 \right)}_{\alpha} r^{\left( 1 \right)}_{\beta} \frac{\partial^2 \mathbf E\left(\mathbf r_0, t \right)}{\partial r_{0 \alpha} r_{0 \beta}} + \frac{\mathbf v^{\left( 2 \right)}}{c} \times \mathbf H\left(\mathbf r_0, t \right) + \frac{\mathbf v^{\left( 1 \right)}}{c} \times \left(\mathbf r^{\left( 1 \right)} \cdot \nabla_0 \right) \mathbf H\left(\mathbf r_0, t \right)\right),
  \end{multline}
  or after some algebra,
  \begin{multline}
    \frac{d}{dt} \left( \mathbf p^{\left(3\right)} - m_e \left(\mathbf r^{\left( 2 \right)} \cdot \nabla_0 \right)\mathbf v^{\left( 1 \right)} - \frac{m_e}{2} r^{\left( 1 \right)}_{\alpha} r^{\left( 1 \right)}_{\beta} \frac{\partial^2 \mathbf v^{\left( 1 \right)}}{\partial r_{0 \alpha} \partial r_{0 \beta}} + m_e v^{\left( 2 \right)}_{\alpha} \nabla_0 r^{\left( 1 \right)}_{\alpha} - m_e \left(\left( \mathbf r^{\left( 1 \right)} \cdot \nabla_0  \right) v^{\left( 1 \right)}_{\alpha} \right) \nabla_0 r^{\left( 1 \right)}_{\alpha}\right) = \\
    = -\frac{m_e}{2} \nabla_0 {\left(\nabla_0 \cdot \left( \mathbf r^{\left( 1 \right)} \left( \mathbf v^{\left( 1 \right)}\right)^2\right) \right)}.
  \end{multline}
  Then a particle after interaction gains the momentum
  \begin{equation}
    \mathbf p^{\left( 3 \right)} \left( {t_\infty}\right) = -\frac{m_e}{2} \nabla_0 \left(\nabla_0 \cdot \int \limits_{-\infty}^{\infty} \mathbf r^{\left( 1 \right)} \left( \mathbf v^{\left( 1 \right)}\right)^2 dt \right),
  \end{equation}
  where it is additionally assumed that $\frac{\partial}{\partial x_i}\mathbf r^{(1)}({t_\infty}) \to 0$. The absorbed angular momentum in the third order of the perturbation theory may be represented as
  \begin{equation}
    \mathbf L^{\left( 3 \right)}  = \mathbf r_0 \times \mathbf p^{\left( 3 \right)} + \mathbf r^{\left( 1 \right)} \times \mathbf p^{\left( 2 \right)} + \mathbf r^{\left( 2 \right)} \times \mathbf p^{\left( 1 \right)}.
  \end{equation}
  This value may appear to be non-zero for certain particles, but for isotropic plasma it again vanishes after averaging
  \begin{equation}
    L^{\left(3\right)}_z (t\to\infty) = -\frac{m_e}{2} \frac{\partial}{\partial \varphi_0} \left( \nabla_0 \cdot \int \limits_{-\infty}^{\infty} \mathbf r^{\left( 1 \right)} \left( \mathbf v^{\left( 1 \right)}\right)^2 dt \right) \xrightarrow[\langle \dots \rangle_{\varphi_0}]{} 0.
  \end{equation}
  %which means that higher orders of perturbation theory are required to describe the total angular momentum gain.
  In the fourth order of the perturbation theory the particle momentum reads
  \begin{multline}
    \frac{d \mathbf p^{\left(4\right)}}{dt} = \frac{m_e c \omega_0}{E_0} a_0 \left( \left(\mathbf r^{\left( 3 \right)} \cdot \nabla_0\right)\mathbf E\left(\mathbf r_0, t \right) + r^{\left( 1 \right)}_{\alpha} r^{\left( 2 \right)}_{\beta} \frac{\partial^2 \mathbf E\left(\mathbf r_0, t \right)}{\partial r_{0 \alpha} \partial r_{0 \beta}} + \frac{1}{6} r^{\left( 1 \right)}_{\alpha} r^{\left( 1 \right)}_{\beta} r^{\left( 1 \right)}_{\gamma} \frac{\partial^3 \mathbf E\left(\mathbf r_0, t \right)}{\partial r_{0 \alpha} \partial r_{0 \beta} \partial r_{0 \gamma}} + \frac{\mathbf v^{\left(3\right)}}{c} \times \mathbf H\left(\mathbf r_0, t \right) + \right.\\
    \left. + \frac{\mathbf v^{\left(2\right)}}{c} \times \left(\mathbf r^{\left(1\right)} \cdot \nabla_0\right)\mathbf H\left(\mathbf r_0, t \right) + \frac{\mathbf v^{\left(1\right)}}{c} \times \left( \left(\mathbf r^{\left(2\right)} \cdot \nabla_0\right)\mathbf H\left(\mathbf r_0, t \right)  + \frac{1}{2} r^{\left( 1 \right)}_{\alpha} r^{\left( 1 \right)}_{\beta} \frac{\partial^2 \mathbf H\left(\mathbf r_0, t \right)}{\partial r_{0 \alpha} \partial r_{0 \beta}} \right) \right).
  \end{multline}

  This expression may be represented as

  \begin{multline} \label{q1}
    \frac{d}{dt} \left( \mathbf p^{\left(4\right)} - m_e \left(\mathbf r^{\left( 3 \right)} \cdot \nabla_0 \right)\mathbf v^{\left( 1 \right)} - m_e r^{\left( 1 \right)}_{\alpha} r^{\left( 2 \right)}_{\beta} \frac{\partial^2 \mathbf v^{\left( 1 \right)}}{\partial r_{0 \alpha} \partial r_{0 \beta}} - \frac{m_e}{6} r^{\left( 1 \right)}_{\alpha} r^{\left( 1 \right)}_{\beta} r^{\left( 1 \right)}_{\gamma} \frac{\partial^3 \mathbf v^{\left( 1 \right)}}{\partial r_{0 \alpha} \partial r_{0 \beta} \partial r_{0 \gamma}} + \right.\\
    \left. + m_e v^{\left( 3 \right)}_{\alpha} \nabla_0 r^{\left( 1 \right)}_{\alpha} - m_e \left(\left( \mathbf r^{\left( 2 \right)} \cdot \nabla_0 \right) v^{\left( 1 \right)}_{\alpha}\right) \nabla_0 r^{\left( 1 \right)}_{\alpha} - \frac{m_e}{2} r^{\left( 1 \right)}_{\alpha} r^{\left( 1 \right)}_{\beta} \frac{\partial^2 v^{\left( 1 \right)}_{\gamma}}{\partial r_{0 \alpha} \partial r_{0 \beta}}\nabla_0 r^{\left( 1 \right)}_{\gamma} + \right.\\
    \left.+ \frac{m_e}{2 c^2} \left(\mathbf v^{\left( 1 \right)}\right)^2 v^{\left( 1 \right)}_\alpha \nabla_0 r^{\left( 1 \right)}_\alpha + m_e \left(\mathbf r^{\left( 2 \right)} \nabla_0\right) \left(\left(\mathbf r^{\left( 1 \right)} \cdot \nabla_0\right) \mathbf v^{\left( 1 \right)}\right)\right) = \\
      = \frac{m_e}{8c^2} \nabla_0 \left( \mathbf v^{\left( 1 \right)}\right)^4 - \frac{m_e}{2} \nabla_0 \left(r^{\left( 1 \right)}_{\alpha} r^{\left( 1 \right)}_{\beta} v^{\left( 1 \right)}_{\gamma} \frac{\partial^2 v^{\left( 1 \right)}_\gamma}{\partial r_{0 \alpha} \partial r_{0 \beta}}\right) - \frac{m_e}{2}\nabla_0 \left( \mathbf v^{\left( 2 \right)}\right)^2 + m_e \frac{d}{dt} \left( \left( \mathbf r^{\left( 2 \right)} \cdot \nabla_0 \right) \mathbf v^{\left( 2 \right)}\right),
  \end{multline}
  where it is taken into account, that $\nabla \cdot \mathbf{r}^{(1)} = 0$, since $\nabla \cdot \mathbf{E}\left(\mathbf r_0, t \right) = 0$.
  The angular momentum, gained by a single particle in the fourth order reads
  \begin{equation}
    \mathbf L^{\left( 4 \right)} = \mathbf r_0 \times \mathbf p^{\left( 4 \right)} + \mathbf r^{\left( 1 \right)} \times \mathbf p^{\left( 3 \right)} + \mathbf r^{\left( 2 \right)} \times \mathbf p^{\left( 2 \right)} + \mathbf r^{\left( 3 \right)} \times \mathbf p^{\left( 1 \right)}.
  \end{equation}
  After averaging its longitudinal component over the azimuthal angle, at ${t\to\infty}$ it becomes
  \begin{equation} \label{<Lz4>}
    \langle L_z^{\left( 4 \right)} ({t_\infty}) \rangle_{\varphi_0} = m_e \langle \left( \mathbf r_0 \times \left(\mathbf r^{\left( 2\right)} \cdot \nabla_0 \right) \mathbf v^{\left( 2\right)} \right)_z + \left(\mathbf r^{\left( 2 \right)} \times \mathbf p^{\left( 2 \right)} \right)_z \rangle_{\varphi_0} \Big|_{{t_\infty}} = \left\langle \left(\mathbf r^{\left( 2\right)} ({t_\infty}) \cdot \nabla_0 \right) L_z^{\left( 2\right)} ({t_\infty}) \right\rangle_{\varphi_0},
  \end{equation}
  which is not zero in general. For the subsequent analysis, obtain here also the expression for the particle energy up to the fourth order of the perturbation theory
  \begin{equation}
    m_e c^2 \gamma = \frac{m_e c^2}{\sqrt{1 - \mathbf v^2/c^2}}\approx m_e c^2 + \frac{m_e}{2} \left(\left(\mathbf v^{\left(1\right)}\right)^2 + 2\left(\mathbf v^{\left(1\right)} \cdot \mathbf v^{\left(2\right)}\right) + 2\left(\mathbf v^{\left(1\right)} \cdot \mathbf v^{\left(3\right)}\right) + \left(\mathbf v^{\left(2\right)}\right)^2\right) + \frac{3 m_e}{8 c^2}\left(\mathbf v^{\left(1\right)}\right)^4.
  \end{equation}
  After the interaction ends, the first non-vanishing contribution to the total kinetic energy gain is also of the fourth order
  \begin{equation}
    m_e c^2 \left(\gamma - 1\right)\vert_{t_\infty}\equiv\varepsilon\vert_{t_\infty} \approx \varepsilon^{(4)} \left( {t_\infty}\right) = \frac{m_e}{2} \left(\mathbf v^{\left(2\right)} \left( {t_\infty} \right) \right)^2.
  \end{equation}
 % where $\varepsilon = m_e c^2 \left(\gamma - 1\right)$ is the kinetic energy of the particle.
\end{widetext}
  Gather the obtained expressions for the momentum, the angular momentum, the average angular momentum and the energy gained by particles in the main non-vanishing order of the perturbation theory on $a_0$:  
  %\bb{should not we also write averaging over $r_0,z_0$? Or say something more general, with the distribution function?}
  \begin{equation} \label{particle_gain_general}
    \begin{cases}
      \begin{aligned}
        & \mathbf p^{(2)} \left( {t_\infty}\right) & = & -\frac{m_e}{2} \nabla_0 \int \limits_{-\infty}^{\infty} \left(\mathbf v^{\left( 1 \right)}{\left(\mathbf r_0, t \right) }\right)^2 dt, \\
        
        & L_z^{(2)} \left( {t_\infty}\right) & = & -\frac{m_e}{2} \frac{\partial}{\partial \varphi_0} \int \limits_{-\infty}^{\infty} \left(\mathbf v^{\left( 1 \right)}{\left(\mathbf r_0, t \right) }\right)^2 dt, \\
        
        & \langle L_z^{(4)} \left( {t_\infty}\right) \rangle_{\varphi_0} & = & \left\langle \left(\mathbf r^{\left( 2\right)} (\mathbf r_0, {t_\infty}) \cdot \nabla_0 \right) L_z^{\left( 2\right) }{\left(\mathbf r_0 , {t_\infty} \right)} \right\rangle_{\varphi_0}, \\
        
        & \varepsilon^{(4)} \left( {t_\infty}\right) & = & \frac{m_e}{2} \left(\mathbf v^{\left(2\right)}{\left(\mathbf r_0 , {t_\infty} \right)} \right)^2.
      \end{aligned}
    \end{cases}
  \end{equation}
  
These expressions represent the central result of the work, they will be used to analyze some certain cases of interaction of the structured light waves with charged particles. Note that up to this moment, the perturbation theory on the wave amplitude $a_0$ was presented, the order of this perturbation theory is designated with the upper-right number in round braces. The perturbation theory on $a_0$ does use Maxwell equations, there are no approximations made for the fields. However, to proceed with certain cases, analytical expressions for the structured light, represented in the next section, are required.
%  It is worth noting, that the next to last term in the right-hand side of $\eqref{q1}$ for longitudinal momentum $p_z$ contains derivative with respect to $z_0$, which may act on a slowly varying envelope.
%  If the envelope is considered to depend on $z$ and $t$ only in combination $t - z$ and the other multiplying factors in the expression for the electromagnetic field does not depend on time, then $\partial/\partial z_0$, acting on the envelope, may be turned into $- \partial/\partial t$, acting on the whole expression.
%  Then the next to last term in the right-hand side of $\eqref{q1}$ integrates to $\varepsilon^{(4)}$.
%  This term is straightforwardly verified to be the leading one in $p_z^{(4)}$ after the interaction is finished in the interaction regime, discussed in the paper.

  \subsection{Approximate description of a structured electromagnetic wave} \label{EM_fields}
%  \textcolor{red}{Specify parameters, such as $w_0$, $\lambda_0$.}
  %In order to develop a theory of interaction, analytic expressions for electromagnetic fields are required.
  As it is explained in Ref. \cite{Thiele2016}, electromagnetic fields in vacuum can be prescribed via boundary conditions for the transverse electric or magnetic field components. %, transverse to the propagation axis of propagation $z$.
  The results in this section are presented for the boundary conditions defined for the magnetic field. Similar results may be obtained using the electric field boundary conditions by substitution $\mathbf E \to -\mathbf H$, $\mathbf H \to \mathbf E$ which follows from the symmetry of Maxwell equations.
%  For the purpose of prescribing electromagnetic fields 
 
  Consider the wave equation for the magnetic field
  \begin{equation}
    \left(\Delta - \frac{1}{c^2}\frac{\partial^2}{\partial t^2}\right) \widetilde{\mathbf H} (\mathbf r, t) = 0,
    \label{initE}
  \end{equation}
  where {$\widetilde{\mathbf E} = \widetilde{\mathbf E}(\mathbf r, t)$ and $\widetilde{\mathbf H} = \widetilde{\mathbf H}(\mathbf r, t)$ are the complex representations of $\mathbf E$ and $\mathbf H$, so that $\mathbf E = \text{Re} \left[\widetilde{\mathbf E} \right]$, $\mathbf H = \text{Re} \left[\widetilde{\mathbf H} \right]$}.
  With $\xi \equiv t - z/c$ as the new time variable the wave equation reads
  \begin{equation}
    \left(\Delta - \frac{2}{c^2}\frac{\partial^2}{\partial \xi \partial z}\right) \widetilde{\mathbf H} {(\mathbf r, \xi)} = 0.
  \end{equation}
%  Introduce the fields in the Fourier space 
%    \begin{equation}
%    \widetilde{\mathbf E} {(\mathbf r,\xi)} = \int \frac{d \omega d \mathbf k}{\left( 2 \pi \right)^4} \mathbf E{(\omega, \mathbf k)} e^{i \left( \omega \xi - \mathbf k \mathbf r \right)},
%  \end{equation}
%  where $\omega$ is the frequency, $\mathbf k = (k_x,k_y,k_z)$ is the wave-vector; here and below all the Fourier-transformed fields are complex and written explicitly with their dependencies, e.g. as $\mathbf E{(\omega, \mathbf k)}$, $\mathbf E{(\omega, \mathbf k_\perp,z)}$ etc.
  Performing the frequency and transversal Fourier transformations
  \begin{equation}
    \widetilde{\mathbf H} {(\mathbf r_\perp, z,\xi)} = \int \frac{d \omega d \mathbf k_\perp}{\left( 2 \pi \right)^3} \mathbf H{(\omega, \mathbf k_\perp,z)} e^{i \left( \omega \xi - \mathbf k_\perp \mathbf r_\perp \right)},
  \end{equation}
  where $\mathbf r_\perp \equiv x \mathbf e_x + y \mathbf e_y$, $\mathbf k_\perp \equiv k_x \mathbf e_x + k_y \mathbf e_y$, obtain that $\mathbf H(\omega, \mathbf k_\perp,z)$ satisfies
  \begin{equation}
    \left( \frac{\partial^2}{\partial z^2} - 2 i \frac{\omega}{c} \frac{\partial}{\partial z} - \mathbf k_\perp^2 \right) \mathbf H(\omega, \mathbf k_\perp,z) = 0.
  \end{equation}
  The two independent solutions are forward and backward propagating waves; the forward propagating wave is defined as
  \begin{equation} \label{forward_solution}
    \mathbf H(\omega, \mathbf k_\perp, z) = \mathbf H(\omega, \mathbf k_\perp, 0) e^{i \omega \left( 1 - \sqrt{1 - \frac{\mathbf k_\perp^2 c^2}{\omega^2}}\right)z/c},
  \end{equation}
  where the condition $\omega^2/c^2 - \mathbf k_\perp^2 > 0$, limiting values of the wave vector for the propagating components, arises.
  The multiplier $\mathbf H(\omega, \mathbf k_\perp, 0)$ is determined with the boundary condition placed at $z = 0$
  \begin{multline}
    \mathbf H(\omega, \mathbf k_\perp, 0) = \int d\xi d\mathbf r_\perp \widetilde{\mathbf H}_\perp \Big|_{z = 0} e^{- i \left( \omega \xi - \mathbf k_\perp \cdot \mathbf r_\perp \right)}, \\ \omega^2/c^2 - \mathbf k_\perp^2 > 0.
    \label{E}
  \end{multline}
  
  The general forward propagating solution reads
  \begin{multline} \label{gen_forward_prop_H}
    \widetilde{\mathbf H} = \\
    \int \limits _{\omega^2/c^2 - \mathbf k_\perp^2 > 0} \frac{d \omega d \mathbf k_\perp}{\left( 2 \pi \right)^3} \mathbf H(\omega, \mathbf k_\perp, 0) e^{i \left( \omega \xi - \mathbf k_\perp \cdot \mathbf r_\perp + \left( \omega/c - k_z \right) z \right)},
  \end{multline}
  where $k_z = \frac{\omega}{c} \sqrt{1 - \frac{\mathbf k_\perp^2 c^2}{\omega^2}}$.
  As it follows from \eqref{gen_forward_prop_H}, the evanescent components of the boundary condition should be dropped, as long as they produce evanescent parts of the solution. Hence the general forward propagating electric and magnetic fields may be written in the form 
  %\bb{where this comes from ? Looks like it should be no $\perp$ in expressions from \eqref{initE} till here, and further in \eqref{longEH} and \eqref{fields_long-perp} say that it is enough to define only perpendicular components.}
  \begin{equation} \label{EH_tilded}
    \begin{aligned}
      {\widetilde{\mathbf E}} & = \int \frac{d \omega d \mathbf k_\perp}{\left( 2 \pi \right)^3} \mathbf E(\omega, \mathbf k_\perp, 0) e^{i \left( \omega \xi - \mathbf k_\perp \cdot \mathbf r_\perp + \left( \omega / c - k_z \right) z \right)}, \\
     { \widetilde{\mathbf H}}  & = \int \frac{d \omega d \mathbf k_\perp}{\left( 2 \pi \right)^3} \mathbf H(\omega, \mathbf k_\perp, 0) e^{i \left( \omega \xi - \mathbf k_\perp \cdot \mathbf r_\perp + \left( \omega / c - k_z \right) z \right)},
    \end{aligned}
  \end{equation}
  where explicit specification of area of integration $\omega^2/c^2 - \mathbf k_\perp^2 > 0$ is omitted.
  
  Transverse and longitudinal components of the electric and magnetic fields at focal points are not independent. %\cite{Thiele2016}.
  As independent parameters of the general forward propagating solution one may choose e.g. the transverse components of the magnetic field, determined with the boundary condition for the transverse components.
  Longitudinal component of the magnetic field and the electric field components may be obtained from Maxwell equations{, which in the new variables read}
  \begin{equation}
  \label{longEH}
    \left\{
      \begin{aligned}
        & \text{div} \widetilde{\mathbf H} = \frac{1}{c}\frac{\partial \widetilde{H_z}}{\partial \xi}, \\
        & \text{rot} \widetilde{\mathbf H} = \frac{1}{c}\left[ \mathbf e_z \frac{\partial \widetilde{\mathbf H}}{\partial \xi}\right] + \frac{1}{c}\frac{\partial \widetilde{\mathbf E}}{\partial \xi},
      \end{aligned}
    \right.
  \end{equation}
  so that
  \begin{equation}
  \label{fields_long-perp}
    \begin{aligned}
      H_z(\omega, \mathbf k_\perp, 0) & = - \frac{\mathbf k_\perp \cdot \mathbf {\mathbf H}_\perp(\omega, \mathbf k_\perp, 0)}{k_z}, \\
      \mathbf E(\omega, \mathbf k_\perp, 0) & = - \frac{c \mathbf k \times {\mathbf H}(\omega, \mathbf k_\perp, 0)}{\omega}.
    \end{aligned}
  \end{equation}
The integration in \eqref{EH_tilded} may be carried out approximately with the expansion of the exponent with the assumption of a slow temporal dependence and slow dependence in the transverse direction
  \begin{eqnarray}
   & \displaystyle e^{i\omega z/c - i \omega \sqrt{1-\frac{\mathbf k_\perp^2 c^2}{\omega^2}}z/c} = e^{i \frac{c \mathbf k_\perp^2}{2 \omega_0}z} \displaystyle \sum\limits_{n,m} \frac{(\omega-\omega_0)^n \mathbf k_\perp^{2m}}{n! m!} \times \nonumber \\
   & \times \frac{\partial^n}{\partial \omega^n} \frac{\partial^m}{\partial \left(\mathbf k_\perp^2\right)^m} e^{i\omega \left(1-\sqrt{1-\frac{\mathbf k_\perp^2 c^2}{\omega^2}} - \frac{c^2 \mathbf k_\perp^2}{2 \omega_0 \omega}\right)z/c}\Bigg\vert_{\substack{\omega=\omega_0 \\ k_\perp = 0}},
      \label{general_expansion}
  \end{eqnarray}
  where $\omega_0$ is the main carrier frequency of the electromagnetic wave.
  Note that according to the paraxial approximation, one has to consider characteristic values to be $k_\perp \sim w_0^{-1}$, $z \sim w_0^2/\lambda_0$, where $w_0$ is the beam waist radius and $\lambda_0$ is the beam wave length, which means that $c \mathbf k_\perp^2 z/\omega_0 \sim 1$ and has to be retained in the exponent in \eqref{general_expansion}. 
  
  For transversely wide beams with the characteristic beam waist radius $w_0$, where the characteristic transverse wave-vector $k_\perp\sim w_0^{-1}$, the paraxial approximation is defined by expansion on $\frac{\mathbf k_\perp^2 c^2}{\omega^2}$, i.e. the order of the paraxial approximation is defined by $m$, see e.g. \cite{Gonzalez2018}. In this framework, the condition $\omega^2/c^2 - \mathbf k_\perp^2$ is neglected in the integration, because it leads to exponentially small corrections. 
  
  For long enough oscillating waves, considered here
  %, the spectrum of the electromagnetic field has a maximum at $\omega_0$,
  and in the case of a slow temporal change of the field amplitude with a characteristic time $\tau$ the condition $\tau \gg 1/\omega_0$ leads to a sharp peak at $\omega = \omega_0$ in the field spectrum.
  Then, the approximate expressions for the electromagnetic field may be obtained \cite{Mora1998} as an expansion of the factor $e^{i\left( \omega/c - k_z \right) z}$ near the peak of the spectrum in $\eqref{EH_tilded}$, the order of the temporal expansion is denoted here by $n$.

  Expansion of the electromagnetic field leads to a series in powers of small parameters related to the inverse beam waist radius $\lambda_0/w_0 \ll 1$ and inverse beam duration $(\omega_0\tau)^{-1} \ll 1$, which may be represented in form
  \begin{equation}
      \mathbf H(\mathbf r, t) = \displaystyle \sum\limits_{\mathbf{n},m} \mathbf H^{\{\mathbf{n},m\}}(\mathbf r, t),
      \label{expansion_general_sum}
  \end{equation}
  where the first (bold) superscript represents the order of the temporal expansion, i.e. the expansion on the inverse beam duration $(\omega_0\tau)^{-1}$ and the second superscript represents the order of the paraxial expansion, i.e. expansion on the inverse beam waist radius $\lambda_0/w_0$. In general, temporal and paraxial expansions may be carried out independently, i.e. leaving only $\mathbf{n}$ or only $m$ summation in \eqref{expansion_general_sum}. For the paraxial expansion only, up to the forth order of $\mathbf k_\perp^2$ one obtains 
  %  \begin{eqnarray}
  %     & e^{i(\omega/c - k_z) z} \approx e^{i \frac{c \mathbf k_\perp^2}{2 \omega_0}z} \times \nonumber \\
  %     & \times \left( 1 - i \frac{c k_\perp^2 (\omega-\omega_0)z}{2\omega \omega_0} +  \frac{c^3 k_\perp^4 z \left( i-\frac{\omega(\omega-\omega_0)^2 z}{c \omega_0^2} \right)}{8\omega^3} + \dots \right), 
  %      \label{exp_expanded_parax}
  % \end{eqnarray}
{
  \begin{eqnarray}
      & e^{i(\omega/c - k_z) z} \approx e^{i \frac{c \mathbf k_\perp^2}{2 \omega}z} \times \nonumber \\
      & \times \left( 1 + i \frac{c^3 \mathbf k_\perp^4}{8 \omega^3} z + i \frac{c^5\mathbf k_\perp^6}{16 \omega^5} z + \dots \right), 
       \label{exp_expanded_parax}
  \end{eqnarray}
  }
which corresponds in the used notations to the paraxial expansion 
  \begin{equation}
      \mathbf H(\mathbf r, t) = \mathbf H^{\{0\}} + \mathbf H^{\{1\}} + \mathbf H^{\{2\}} + \dots,
  \end{equation}
where only the second index in \eqref{expansion_general_sum} is left.

In the same way, the expansion near $\omega_0$, i.e.
   \begin{multline}
       e^{i(\omega/c - k_z) z} \approx e^{i\omega_0 z/c - i \omega_0 \sqrt{1-\frac{\mathbf k_\perp^2} {\omega_0^2}}z/c} \times  \\
       \times \left( 1 + i\left( 1 -\frac{\omega_0}{\sqrt{\omega_0^2- c^2 \mathbf k_\perp^2}}\right)\frac{z\left( \omega-\omega_0\right)}{c} + \right. \label{exp_expanded_temp} \\
       +  \left( \frac{i c^3 \mathbf k_\perp^2}{z\left( \omega_0^2-c^2 \mathbf k_\perp^2 \right)^{3/2}} - \right.  \\
       - \left. \left. \left(  1 -\frac{\omega_0}{\sqrt{\omega_0^2-c^2 \mathbf k_\perp^2}} \right)^2 \right)\frac{z^2(\omega-\omega_0)^2}{2c^2}+ \dots \right), 
  \end{multline}
provides the temporal expansion 
  \begin{equation}
      \mathbf H(\mathbf r, t) = \mathbf H^{\{\mathbf{0}\}} + \mathbf H^{\{\mathbf 1\}} + \mathbf H^{\{\mathbf 2\}} + \dots.
  \end{equation}
Similar expressions may be presented for the electric field components.

%  Considering expansion 
  % \begin{equation}
  %     k_z = \omega/c - \frac{c \mathbf k_\perp^2}{2 \omega} - \frac{c^3 \mathbf k_\perp^4}{8 \omega^3} + \dots,
  % \end{equation}
  To take into account both the temporal and paraxial corrections in a few low orders, the exponent $e^{i(\omega/c - k_z) z}$ in \eqref{EH_tilded} may be expanded in powers of $\lambda_0/w_0 \ll 1$ and $(\omega_0\tau)^{-1} \ll 1$ as
  \begin{multline} \label{exp expanded}
      e^{i(\omega/c - k_z) z} \approx \\  e^{i \frac{c \mathbf k_\perp^2}{2 \omega_0}z} \left( 1 + i \frac{c^3 \mathbf k_\perp^4}{8 \omega_0^3}z - i(\omega - \omega_0) \frac{c \mathbf k_\perp^2}{2\omega_0^2}z + \dots\right).
  \end{multline}
  The first term corresponds to the lowest order paraxial approximation and the constant envelope, the second relates to the first non-vanishing correction to the lowest order paraxial approximation with the constant envelope and the third one corresponds to the temporal corrections taking into account the time variations of the envelope in the lowest order paraxial approximation.
  From $\eqref{fields_long-perp}$ and $\eqref{exp expanded}$ one can see, that the the transverse components of electromagnetic field are expanded in even powers of $\lambda_0/w_0$ and longitudinal in odd powers of $\lambda_0/w_0$.
    
    {Having in mind a possibility of a general analysis, consider the electromagnetic fields as a linear combination of modes
  \begin{equation}
    \begin{cases}
    \mathbf{H}=\sum\limits_i \mathbf{H}^{(i)}, \\
    \mathbf{E}=\sum\limits_i \mathbf{E}^{(i)}, 
    \label{linear_sum}
 \end{cases}
 \end{equation}
  which will be later specified for several certain examples.}
  The boundary condition for transverse components in the $(x,y)$ plane is defined for the magnetic field as
  \begin{equation}
    \mathbf H^{(i)}_\perp \Big|_{z = 0} = \mathbf H^{(i)}_{0\perp}\left( \mathbf r_\perp, t \right),
  \end{equation}
  {where $z = 0$ is the focal plane. Consider here sufficiently mildly focused beams, so that for the chosen $\mathbf H^{(i)}_{0\perp}\left( \mathbf r_\perp, t \right)$ no evanescent components appear.}
  
  {For the main goal of this work, it is enough to consider the light-particle OAM absorption in the first non-vanishing order of the interaction parameters, defined according to the expansion \eqref{expansion_general_sum}.  So, the paraxial approximation $\lambda_0/w_0 \ll 1$ and the approximation of a slowly varying temporal envelope $\left(\tau \omega_0\right)^{-1} \ll 1$, are assumed.} %\cite{Mora1998}.
  %
%  The following sections of the paper consider time measured in units of $\omega_0^{-1}$, length in units of $k_0^{-1} = \frac{\lambda_0}{2 \pi}$, mass in units of electron mass $m$ and electric and magnetic fields in units of a reference electric field $E_0$.
%  Thus, the equations of motion $\eqref{dim_eqs_motion}$ take form
%
%  \begin{equation}
%    \frac{d\mathbf p}{dt} = a_0 \left( \mathbf E + \left[ \mathbf v \mathbf H \right] \right),
%  \end{equation}
%  where $a_0 = \frac{eE_0}{m\omega_0 c}$ is the dimensionless field amplitude.
%
  {The boundary conditions in the focal plane are chosen for convenience in the complex form, so that $H^{(i)}_{x, y} = \text{Re} \left[ \widetilde{H}^{(i)}_{x, y} \right]$:
  \begin{equation} \label{boundary_condition}
    \begin{cases}
      \begin{aligned}
        \widetilde{H}^{(i)}_x \Big|_{z = 0} & = E_0 g \left( t \right) e^{i \omega_0 t} \mathbfcal{H}_{0 \perp}^{(i)} \left( \mathbf r_\perp\right), \\
        \widetilde{H}^{(i)}_y \Big|_{z = 0} & = -i \sigma^{(i)} \widetilde{H}^{(i)}_x \Big|_{z = 0},
      \end{aligned}
    \end{cases}
  \end{equation}
  where $g \left( t\right)$ is a common slow temporal envelope, $E_0$ is the common field amplitude and $\mathbfcal{H}_{0 \perp}^{(i)} \left( \mathbf r_\perp\right) = \mathcal H_{0x}^{(i)}( \mathbf r_\perp) \mathbf{ e_x}+ \mathcal H_{0y}^{(i)} ( \mathbf r_\perp) \mathbf{ e_y}$ is an arbitrary function.
  The used setup \eqref{boundary_condition} represents a superposition of co-propagating along $z$ axis circularly ($\sigma^{(i)} = \pm 1$) or linearly ($\sigma^{(i)} = 0$) polarized beams.

  The function $\mathbfcal{H}_{0 \perp}^{(i)} \left( \mathbf r_\perp\right)$ may be expressed in terms of the eigenfunctions of the paraxial wave equation $u_{pl}$, which are defined as \cite{Allen1992} 
  %$a^{(i)}_{pl}$ are the dimensionless coefficients, $U_{pl} \left( \mathbf r_\perp/w_0 \right)$ corresponds to a Laguerre-Gaussian mode in the focal plane $U_{pl} \left( \mathbf r_\perp/w_0 \right)=u_{pl}(\mathtt{r},\phi,\mathtt{z}=0)$, where $\mathtt{r}=r_\perp/w_0$ and $\mathtt{z}=z/z_R$ with $z_R = \frac{\pi w_0^2}{\lambda_0}$ being the Rayleigh length, are the dimensionless transverse radius and longitudinal coordinate, $u_{pl}$ are defined as \cite{Allen1992}
  \begin{multline} \label{u_pl}
    u_{pl} \left( \mathtt{r}, \varphi, \mathtt{z} \right) = C_{pl} \frac{1}{\mathtt{w} \left( \mathtt{z} \right)} \times \\
    \left(\frac{\mathtt{r}\sqrt{2}}{\mathtt{w} \left( \mathtt{z} \right)}\right) ^ {\left| l \right|} \exp{\left(-\frac{\mathtt{r}^{2}}{\mathtt{w}^{2} \left( \mathtt{z} \right)} \right)} L_{p}^{\left| l \right|}\left(\frac{2 \mathtt{r}^{2}}{\mathtt{w}^{2} \left( \mathtt{z}\right)} \right) \times \\
    \exp {\left( - i l \varphi - i \frac{\mathtt{r}^{2} \mathtt{z}}{\mathtt{w}^2 \left( \mathtt{z}\right)}
    + i \left( 2p + \left| l \right| + 1\right) \tan^{-1} \left(\mathtt{z} \right)\right)},
  \end{multline}
  where $\mathtt{r}=r_\perp/w_0$ and $\mathtt{z}=z/z_R$ with $z_R = \frac{\pi w_0^2}{\lambda_0}$ being the Rayleigh length, are the dimensionless transverse radius and longitudinal coordinate, $C_{pl} = \sqrt{\frac{2}{\pi} \frac{p !}{\left( p + \left| l \right|\right)!}}$ is the normalization constant, $\mathtt{w}\left( \mathtt{z} \right) = \sqrt{1 + \mathtt{z}^2}$,  $L_{p}^{\left| l \right|}$ is the generalized Laguerre polynomial.}
  
  The basis for the boundary conditions is formed by the functions $u_{pl} (\mathtt{r},\varphi,\mathtt{z}=0) \equiv U_{pl}(\mathtt{r},\varphi)$.
  Then the expansion of $\mathbfcal{H}_{0 \perp}^{(i)} \left( \mathbf r_\perp\right)$ reads
  \begin{equation}
    \mathcal H_{0\alpha}^{(i)} \left( r_\perp, \varphi\right) = \sum \limits_{p = 0}^\infty \sum \limits_{l = -\infty}^\infty a_{pl \alpha}^{(i)} U_{pl} \left( \mathtt{r}, \varphi\right),
  \end{equation}
  where
  \begin{equation}
    a_{pl \alpha}^{(i)} = \int\mathtt{r}d\mathtt{r} d\varphi\mathcal H_{0\alpha}^{(i)} \left( r_\perp, \varphi\right)  U_{pl}^* \left( \mathtt{r}, \varphi\right), \label{a_plalpha}
  \end{equation}
and $\alpha = x$ or $y$.

Substituting the boundary condition $\eqref{boundary_condition}$, expanded in terms of $U_{pl}$, to \eqref{E}, integrating over $\xi$, introducing the angle between $\mathbf k_\perp$ and $x$-axis, such that $\mathbf k_\perp \cdot \mathbf r_\perp = k_\perp r_\perp \cos(\theta - \varphi)$, and integrating over the azimuthal angle $\varphi$, one obtains \eqref{Bessel_in_Ealpha}, using the representation of the Bessel function \eqref{Bessel_representation}. There, $g_{\omega} = \int d\xi g\left( \xi \right) e^{-i \omega \xi}$ is the Fourier transform of the envelope. 
%Then, using \eqref{ryzhik_identity}
%integrating over $\xi$ and azimuthal angle $\varphi$ results in the Bessel function $2\pi i^s J_s(z)=2\pi i^{|s|} J_{|s|}(z)=\int_0^{2\pi} d\phi \exp [i s \phi + i z \cos \phi]$ with integer $s$
%\begin{equation}
%     E_\alpha(\omega, \mathbf k_\perp, z=0) = 2 \pi g_{\omega - 1}   \sum \limits_{p, l} a_{pl\alpha} i^{|l|} e^{- i l \theta} \int r dr U_{pl} (r/w_0, 0)  J_{|l|}(|\mathbf k_\perp| r),
% \end{equation}
% where $g_{\omega} = \int d\xi g\left( \xi \right) e^{-i \omega \xi}$ is the Fourier transform of the envelope and $\theta$ is the angle between $\mathbf k_\perp$ and $x$-axis, such that $\mathbf k_\perp \cdot \mathbf r_\perp = |\mathbf k_\perp| r \cos(\theta - \varphi)$.
% \bb{In $U_{pl}$ there is no $|l|$ in $e^{il\varphi}$ -- is it correctly counted in \eqref{E_alpha-from-boundary}}
Then, integrating over $\mathtt r$ and using the identity \eqref{ryzhik_identity}, one obtains
  \begin{multline}
    H_\alpha^{(i)}(\omega, \mathbf k_\perp, z=0) = \\
    =
    E_0 g_{\omega - \omega_0} \pi w_0^2 \sum \limits_{p, l} a_{pl \alpha}^{(i)} i^{2 p + |l|} U_{pl} \left( \frac{ k_\perp w_0}{2}, \theta\right).
    \label{E_alpha-from-boundary}
  \end{multline}
  The lowest order approximation of the solution then takes form
  \begin{equation}
    \widetilde H_{\alpha}^{\{\mathbf 0, 0\}} = E_0 g \left( \xi \right) \sum \limits_i \sum \limits_{p, l} a_{pl \alpha}^{(i)} u_{pl} \left( \mathtt{r}, \varphi, \mathtt{z}\right) e^{i \omega_0 \xi},
  \end{equation}
  where the first (bold) index in $\widetilde H_{\alpha}^{\{\mathbf 0, 0\}}$ corresponds to the order of expansion on $(\omega_0\tau)^{-1}$ and the second to the order of expansion on $\lambda_0/w_0$.
  Introducing
  \begin{equation} \label{H_alpha_00}
    \mathcal H_{\alpha}^{\{\mathbf 0,0\}} \left( \mathbf r \right) = \sum \limits_i \sum \limits_{p, l} a_{pl \alpha}^{(i)} u_{pl} \left( \mathtt{r}, \varphi, \mathtt{z}\right),
  \end{equation}
  such that $\mathcal H_{\alpha}^{\{\mathbf 0,0\}} \left( \mathbf r \right) \Big|_{z = 0} = \mathcal H_{0 \alpha} \left( \mathbf r_\perp \right) \equiv \sum \limits_i \mathcal H_{0 \alpha}^{(i)} \left( \mathbf r_\perp \right)$,
  the lowest order approximation may be represented in the form
  \begin{equation}
    \widetilde{\mathbf H}_{\perp}^{\{\mathbf 0,0\}} = E_0 g \left( \xi \right) \mathbfcal{H}_{\perp}^{\{\mathbf 0,0\}} \left( \mathbf r \right) e^{i \omega_0 \xi}.
  \end{equation}
  The first correction corresponding to the finite pulse  duration
  \begin{equation}
    \widetilde{\mathbf H}_{\perp}^{\{\mathbf 1,0\}} = \frac{E_0}{\omega_0} g^\prime \left( \xi \right) i z \frac{\partial \mathbfcal{H}_{\perp}^{\{\mathbf 0,0\}}}{\partial z} e^{i \omega_0 \xi}, \label{E10}
  \end{equation}
  As one can see, this correction term is zero at the focal point, which is in agreement with the boundary condition $\eqref{boundary_condition}$.

  The first non-vanishing contribution to the longitudinal component of the magnetic field is
  \begin{equation}
      \widetilde{\mathbf H}_{\perp}^{\{\mathbf 0,1\}} = E_0 g \left( \xi \right) \mathcal{H}_z^{\{\mathbf 0,1\}} \left( \mathbf r \right) e^{i \omega_0 \xi},
  \end{equation}
  where $\mathcal{H}_z^{\{\mathbf 0,1\}} = - i \frac{c}{\omega_0}\nabla_\perp \cdot \mathbfcal{H}_{\perp}^{\{\mathbf 0, 0\}}$.
  The first non-vanishing correction to the transverse components corresponding to the paraxial approximation is
  \begin{equation}
    \widetilde{\mathbf H}_{\perp}^{\{\mathbf 0,2\}} = \frac{c E_0}{\omega_0} g \left( \xi \right) \frac{z}{2i} \frac{\partial^2 \mathbfcal{H}_{\perp}^{\{\mathbf 0,0\}}}{\partial z^2} e^{i \omega_0 \xi}, \label{E02}
  \end{equation}
  which is zero at the focal point.
  The next corrections and expressions for the electric field components are presented in Appendix \ref{Field_components}.

  % In regular CGS units lowest order approximation takes form
  % \begin{equation}
  %   \widetilde{\mathbf E}_{\perp}^{\{0, 0\}} = E_0 g \left( \xi \right) \mathbfcal{E}_{\perp}^{\{0, 0\}} \left( \mathbf r_\perp \right) e^{i \omega_0 \xi}.
  % \end{equation}
  
  % The first correction corresponding to finite duration in CGS units
  % \begin{equation}
  %   \widetilde{\mathbf E}_{\perp}^{\{1, 0\}} = \frac{E_0}{\omega_0} g^\prime \left( \xi \right) i z \frac{\partial \mathbfcal{E}_{\perp}^{\{0, 0\}}}{\partial z} e^{i \omega_0 \xi}, \\
  % \end{equation}

  % The first (non-vanishing) correction corresponding to paraxial approximation in CGS units
  % \begin{equation}
  %   \widetilde{\mathbf E}_{\perp}^{\{0, 2\}} = \frac{E_0 c}{\omega_0} g \left( \xi \right) \frac{z}{2i} \frac{\partial^2 \mathbfcal{E}_{\perp}^{\{0, 0\}}}{\partial z^2} e^{i \omega_0 \xi}.
  % \end{equation}

{In what follows, the general approximate expressions are presented for the momentum, energy, and the orbital angular momentum (OAM), absorbed by uniformly ditributed electrons from the mildly focused laser pulse of a sub-relativistic intensity. These general expressions are derived in the frames of the perturbation theory developed on the field amplitude $a_0$, the paraxial parameter $\lambda_0/w_0$ and the adiabatic parameter of the laser pulse $(\omega_0 \tau)^{-1}$, aiming the accurate consideration of the main non-vanishing contributions. The calculations show, that for the OAM absorption, it is enough to consider the fourth order on $a_0$, the first order on $(\omega_0 \tau)^{-1}$ (the first correction to the main order) and the main order of the paraxial approximation.}

% \bb{Based on the obtained general results, the special cases of the laser beams are considered, namely:
% \begin{enumerate}
% \item the circularly polarized beam with a single Laguerre-Gaussian mode,
% \item the linearly polarized beam with a single Laguerre-Gaussian mode,
% \item the beam being a superposition of two linearly polarized Laguerre-Gaussian modes,
% \item ...
% \end{enumerate}}

  \subsection{Analytical estimates for the energy, momentum and angular momentum gained by a particle in a structured wave}
  
  %In order to accurately describe the angular momentum gain by electrons, one has to consider corrections to the lowest order slowly varying envelope approximation solutions of Maxwell equations.
  As it is shown below, the first order temporal correction provides a contribution to the average angular momentum gain by electrons of the same order as the constant amplitude approximation, and hence should be considered in the calculations to obtain a correct expression for the average angular momentum gain.
%  If not stated otherwise, the lowest order of paraxial and slowly varying envelope approximations is considered.
  Magnetic field accounting for the first order {temporal} correction may be represented, according to \eqref{exp_expanded_temp} and \eqref{E10} as
  \begin{multline} %\label{H_expanded}
    \mathbf H = E_0 \text{Re} \left\{ \left( g \left( t - z / c\right) \mathbfcal{H}^{\{\mathbf{0}\}}\left( \mathbf r \right) + \right. \right. \\
    \left. \left. \frac{1}{\omega_0} g^\prime \left( t - z / c \right) \mathbfcal{H}^{\{\mathbf{1}\}} \left( \mathbf r \right) \right) e^{i \omega_0 \left( t - z / c \right)} \right\}, \label{H}
  \end{multline}
  where both $\mathbfcal{H}^{\{\mathbf{0}\}}$ and $\mathbfcal{H}^{\{\mathbf{1}\}}$ are calculated with a desired accuracy on the paraxial parameter $\lambda_0 / w_0$.
  The derivative of the envelope $g^\prime \left( t - z / c \right)$ may be estimated as $g^\prime \sim \frac{g}{\omega_0 \tau} \ll g$, which makes this term a small correction to the leading one.
  In addition, this condition means that $g(t - z/c)$ is a smooth function, compared to $e^{i\omega_0(t - z/c)}$, and we also consider $g^\prime(t - z/c)$ to be a smooth function as well.
  
  The spatial amplitude $\mathbfcal{H}^{\{\mathbf{0}\}}$ is defined from the boundary condition \eqref{boundary_condition}.
  The lowest order paraxial approximation for the transverse components is \eqref{H_alpha_00} and the lowest order paraxial approximation for the transverse components of $\mathbfcal{H}^{\{\mathbf{1}\}}(\mathbf r)$ is defined as \eqref{E10}.
  %obtained straightforwardly
  %\begin{equation}
  %  \begin{cases}
  %    \begin{aligned}
  %      \mathcal H_{x}^{\{\mathbf 0, 0\}}(\mathbf r) & = \sum \limits_{i} \sum \limits_{p, l} a^{(i)}_{pl} u_{pl} \left(r, \varphi, z \right), \\
  %      \mathcal H_{y}^{\{\mathbf 0, 0\}}(\mathbf r) & = -i \sum \limits_{i} \sigma^{(i)} \sum \limits_{p, l} a^{(i)}_{pl} u_{pl} \left(r, \varphi, z \right),
  %    \end{aligned}
  %  \end{cases}
  %\end{equation}
  %where the first upper superscript is related as before to the temporal corrections, while the second upper superscript is introduced to represent here and below the order of the paraxial approximation.
  
  %The lowest order paraxial approximation for the transverse components of $\mathbfcal{H}^{\{\mathbf{1}\}}(\mathbf r)$ in \eqref{H} is defined as
  %\begin{equation}
  %    \mathbfcal{H}_{\perp}^{\{\mathbf 1, 0\}} = i z \frac{\partial \mathbfcal{H}_{\perp}^{\{\mathbf 0, 0\}}}{\partial z}.
  %\end{equation}
  %As one can see, this correction term is zero at the focal point, which is in agreement with the boundary condition $\eqref{boundary_condition}$.
  %\bb{\textbf{There is no $g'$ in \eqref{boundary_condition}!}}.

  The electric field may be written in the similar form as \eqref{H} 
  \begin{multline} \label{E_expanded}
    \mathbf E = E_0 \text{Re} \left\{ \left( g \left( t - z / c\right) \mathbfcal{E}^{\{\mathbf{0}\}} \left( \mathbf r \right) + \right. \right. \\
    \left. \left. \frac{1}{\omega_0} g^\prime \left( t - z / c \right) \mathbfcal{E}^{\{\mathbf{1}\}} \left( \mathbf r \right) \right) e^{i \omega_0 \left( t - z / c \right)} \right\},
  \end{multline}
  where $\mathbfcal{E}^{\{\mathbf{0}\}}$ and $\mathbfcal{E}^{\{\mathbf{1}\}}$ are the spatial amplitudes, defined from the amplitudes of the magnetic field, which in a few lowest orders of paraxial approximation are given by
  \begin{equation}
      \begin{aligned}
        \mathcal E_{x}^{\{\mathbf 0, 0\}}(\mathbf r) & = \mathcal H_{y}^{\{\mathbf 0, 0\}}(\mathbf r) \\
        \mathcal E_{y}^{\{\mathbf 0, 0\}}(\mathbf r) & = - \mathcal H_{x}^{\{\mathbf 0, 0\}}(\mathbf r), \\
        \mathbfcal{E}_{\perp}^{\{\mathbf 1, 0\}} (\mathbf r) & = i z \frac{\partial \mathbfcal{E}_{\perp}^{\{\mathbf 0, 0\}} (\mathbf r)}{\partial z},
      \end{aligned}
  \end{equation}
  see more details in the Appendix.

%\bb{Explain notations: if $\nabla$ is always the gradient, may say this. If it is somewhere something else, should be corrected to be clear. Say also what means directional derivative and grad of div.}  

  Substituting $\eqref{E_expanded}$ into $\eqref{particle_gain_general}$, in the main order of the slowly varying envelope approximation one obtains
  \begin{equation} \label{p_ekin_particular}
    \begin{aligned}
      \mathbf p^{(2)} \left( {t_\infty}\right) & = - m_e c^2 \frac{a_0^2}{4} \nabla_0 \left| \mathbfcal{E}^{\{\mathbf 0\}} \right|^2 \tau_{int}, \\
      L_z^{(2)} \left( {t_\infty}\right) & = - m_e c^2 \frac{a_0^2}{4} \frac{\partial}{\partial \varphi_0} \left| \mathbfcal{E}^{\{\mathbf 0\}} \right|^2 \tau_{int}, \\
      \varepsilon^{(4)} \left( {t_\infty}\right) & = m_e c^4 \frac{a_0^4}{32} \left(\nabla_0 \left| \mathbfcal{E}^{\{\mathbf 0\}} \right|^2 \right)^2 \tau_{int}^2,
   \end{aligned}
  \end{equation}
  where $\tau_{int} \equiv \int \limits_{-\infty}^\infty g^2\left( t \right) dt \sim \tau$ is the characteristic time of the interaction.

  In order to calculate the average gained angular momentum in the fourth order, one has to calculate $\mathbf r^{(2)}$ and $L_z^{(2)}$ after the interaction with the laser.
  %Angular momentum $L_z^{(2)}$ is calculated in $\eqref{p_ekin_particular}$, thus only $\mathbf r^{(2)}$ remains to be obtained.
  The lowest order of $\mathbf r^{(2)}$ arises from the second term in $\eqref{r2}$.
  At infinite time it diverges into infinity, and should be regularized.
  The regularization may be obtained as setting the last moment of time to be $T$, assuming ${T\to\infty}$ at the end of calculations.
  Substituting the first term of $\eqref{E_expanded}$ and evaluating $\mathbf r^{(2)}$ at high values of time $T$, one obtains
  \begin{equation}
      \mathbf r^{(2)}(T) \approx - c^2 \frac{a_0^2}{4} \nabla_{0} \left| \mathbfcal{E}^{\{\mathbf 0\}}\right|^2 \int \limits_{-\infty}^{T} dt \int \limits_{-\infty}^t g^2(t^\prime) dt^\prime.
  \end{equation}

  The corresponding contribution to the average angular momentum is
  \begin{multline}
      \langle L_z^{(4)} \left( T\right) \rangle_{\varphi_0} \approx m_e c^4 \frac{a_0^4}{32} \Bigl \langle \frac{\partial}{\partial \varphi_0} \left( \nabla_0 \left| \mathbfcal{E}^{\{\mathbf 0\}} \right|^2 \right)^2 \Bigl \rangle_{\varphi_0} \times \\
      \tau_{int} \int \limits_{-\infty}^{T} dt \int \limits_{-\infty}^t g^2(t^\prime) dt^\prime = 0,
  \end{multline}
  which is a partial derivative with respect to $\varphi_0$ and vanishes being averaged with the isotropic distribution function.
  This expression does not depend on $T$ and the limit ${T\to\infty}$ may be applied.

  The next contribution to average angular momentum gain results from both terms in $\mathbf r^{(2)}$ in $\eqref{r2}$.
  The first one may be evaluated in the lowest order.
  In the second term one has to consider correction to the slowly varying envelope approximation.
  
  The first term in the lowest order expansion in powers of $(\omega_0 \tau)^{-1}$ reads
  \begin{multline}
      \int \limits_{-\infty}^\infty dt^\prime \left(\mathbf r^{\left( 1 \right)} (t^\prime) \cdot \nabla_0\right)\mathbf v^{\left( 1 \right)} (t^\prime) \approx \\
      \approx \frac{c^2}{\omega_0} \frac{a_0^2}{2} \text{Re}\left(i \left(\mathbfcal{E}^{\{\mathbf 0\} *} \cdot \nabla_0\right) \mathbfcal{E}^{\{\mathbf 0\}} + \mathcal{E}^{\{\mathbf 0\}}_z \mathbfcal{E}^{\{\mathbf 0\} *}\right) \tau_{int} \approx \\
      \approx -\frac{c^2}{\omega_0} \frac{a_0^2}{2} \frac{\partial}{\partial x_{j0}}\text{Re}\left(i \mathcal{E}_j^{\{\mathbf 0\}} \mathbfcal{E}^{\{\mathbf 0\} *}\right) \tau_{int},
      \label{r2_1}
      \end{multline}
  where $\left(\nabla \cdot \mathbfcal{E}^{\{\mathbf 0\}}\right) \approx i \mathcal{E}_z^{\{\mathbf 0\}}$ in the lowest order was used.

  The second term consists of two parts: the first one arises from the action of $\nabla_0$ on $g(t - z_0/c)$, and the second one from the correction to the slowly varying envelope approximation.
  This results in the two following contributions to $\mathbf r^{(2)}$ correspondingly
  \begin{equation}
      c \frac{a_0^2}{4}\left| \mathbfcal{E}^{\{\mathbf 0\}}\right|^2 \tau_{int} \mathbf e_z
      \label{r2_2}
  \end{equation}
  and
  \begin{equation}
  - c \frac{a_0^2}{8} \nabla_0 \text{Re}\left( \mathbfcal{E}^{\{\mathbf 0\}} \mathbfcal{E}^{\{\mathbf 1\}*}\right) \tau_{int}.
  \label{r2_3}
  \end{equation}
  Collecting the three terms \eqref{r2_1}, \eqref{r2_2} and \eqref{r2_3} together, substituting to the expression \eqref{<Lz4>} for $\langle L_z^{(4)} \rangle_{\varphi_0}$  and integrating the first two of them by parts inside the averaging one obtains
  \begin{widetext}
    \begin{multline} \label{Lz_particular_4_order}
      \langle L_z^{(4)} \left( {t_\infty}\right) \rangle_{\varphi_0} = \frac{m_e c^4}{\omega_0} \frac{a_0^4}{8} \Bigl \langle \left\{ \left[ \frac{1}{4} \left( \nabla_0 \text{Re}\left( \mathbfcal{E}^{\{\mathbf 0\}} \mathbfcal{E}^{\{1\} *}\right) \cdot \nabla_0 \right) + \frac{\omega_0}{2 c} \frac{\partial \left| \mathbfcal{E}^{\{\mathbf 0\}} \right|^2}{\partial z_0}\right] \frac{\partial \left| \mathbfcal{E}^{\{\mathbf 0\}} \right|^2}{\partial \varphi_0} + \right.\\
      \left. + \frac{\partial}{\partial x_{k0}} \left( \frac{\partial}{\partial x_{j0}} \text{Re}\left( i \mathcal E_j^{\{\mathbf 0\}} \mathcal E_k^{\{\mathbf 0\} *} \right) \frac{\partial \left| \mathbfcal{E}^{\{\mathbf 0\}} \right|^2}{\partial \varphi_0}\right) \right\} \Bigl \rangle_{\varphi_0} \tau_{int}^2.
    \end{multline}
  \end{widetext}

  For a  Laguerre-Gaussian beam with a characteristic transverse size $w_0$ and the longitudinal size $z_R \sim w_0^2/\lambda_0$, the expressions $\eqref{p_ekin_particular}$ and $\eqref{Lz_particular_4_order}$ may be estimated in the main orders as
  \begin{equation} \label{estimations}
    \begin{aligned}
      \big|\mathbf p_\perp^{(2)} \left( {t_\infty}\right) \big| & \sim m_e c^2 \frac{a_0^2 \tau_{int}}{w_0}, \\
      p_z^{(2)} \left( {t_\infty}\right) & \sim \frac{m_e c^3}{\omega_0} \frac{a_0^2 \tau_{int}}{w_0^2}, \\
      L_z^{(2)} \left( {t_\infty}\right) & \sim m_e c^2 a_0^2 \tau_{int}, \\
      \varepsilon^{(4)} \left( {t_\infty}\right) & \sim m_e c^4 \frac{a_0^4}{w_0^2} \tau_{int}^2, \\
      \langle L_z^{(4)} \left( {t_\infty}\right) \rangle_{\varphi_0} & \sim \frac{m_e c^4}{\omega_0} \frac{a_0^4 \tau_{int}^2}{w_0^2}.
   \end{aligned}
  \end{equation}
  Note, that the transverse and the longitudinal characteristic sizes are different, which makes estimations for transverse and longitudinal momenta also to be different.

The higher orders of the perturbation theory has the $\sim\tau_{int}^2$ dependence, meaning formally that for later interaction times the energy and OAM transfer becomes more efficient. These estimates however actually require that the forth order terms are less than the second order ones, or, using the expressions for $\langle L_z^{(4)}\left( {t_\infty}\right) \rangle_{\varphi_0} $ and for $L_z^{(2)} \left( {t_\infty}\right)$, that $a_0^2 \tau_{int}/w_0^2 \ll \omega_0 / c^2$. This means that for long pulses, for which $\tau_{int} \gtrsim \omega_0 w_0^2 / c^2 a_0^2$, after the moment of time $t \sim \omega_0 w_0^2 / c^2 a_0^2$, the perturbation theory fails. The obtained limitation for the interaction time may be rewritten as $\omega_0\tau_{int} \ll \left( w_0/l_{osc}\right)^2$, where $l_{osc} = a_0 \lambda_0$ is the characteristic oscillation amplitude of the electron in the wave.
%As a result, the perturbation theory fails when the oscillation amplitude $l_{osc}$ appears to be comparable to the beam waist radius $w_0$.
%This condition means that the electron's oscillation amplitude 

Comparing the second and the forth-order expressions for the transferred momentum in \eqref{estimations}, one can see, that $l_{osc}^2 \omega_0 \tau_{int} / w_0^2$ is the actual perturbation theory parameter for the expansion of gained electron values, which counts for both the paraxial approximation and the slow-varying envelope approximation, and means, that the field amplitude $a_0$ may be actually not too small for the applicability of the obtained results. This is confirmed by the examples presented in the next section.  

%For the maximum values of the transferred orbital momentum and the rate of transfer, one may substitute the interaction time by its upper limit, which results in the following estimates
%   \begin{equation} \label{estimation_L_max}
%   \begin{aligned}
%       L_z^{(max)} \left( {t_\infty}\right)  \sim m_e w_0^2 \omega_0;\\
%       L_z^{(max)}/\tau_{int}  \sim m_e c^2 a_0^2 .
% \end{aligned}
%   \end{equation}
% The first actually corresponds to the electrons rotating with the frequency of the driving wave at the distance $w_0$ around $z$ axis, which requires that it has velocity greater than the light velocity, so should be limited as
%   \begin{equation} \label{estimation_L_max_limited}
%       L_z^{(max)} \left( {t_\infty}\right)  \lesssim m_e c w_0,
%   \end{equation}
%   but on the other hand, this overestimation means that the interaction time in estimations \eqref{estimations} is limited.
  According to the arguments provided just above, although the perturbation theory was developed under the formal assumption of a low amplitude $a_0$, a moderate intensity regime of the interaction $a_0 \sim 1$ is considered in the following part of the paper. %As will be demonstrated further, perturbation theory provides a qualitative agreement between the numerical simulations and perturbation theory even at moderate regime $a_0 \sim 1$.
  At field amplitudes $a_0 \sim 1$ the following estimates may be obtained 
  \begin{equation} \label{estimations a0=1}
    \begin{aligned}
      p_z^{(2)} \left( {t_\infty}\right) & \sim \frac{m_e c^3}{\omega_0} \frac{\tau_{int}}{w_0^2}, \\
      \varepsilon^{(4)} \left( {t_\infty}\right) & \sim m_e c^4 \frac{1}{w_0^2} \tau_{int}^2, \\
      (p_z - \varepsilon / c) \Big|_{{t_\infty}} & \sim \frac{m_e c^3}{\omega_0} \frac{\tau_{int}}{w_0^2},
   \end{aligned}
  \end{equation}
  the first two are the consequence of $\eqref{estimations}$ and the last one does not depend on perturbation theory and may be obtained from the analysis of integrals of motion. As it was shown in Ref. \cite{dmitriev2022effect}, the growth rate of $(p_z - \varepsilon / c)$ is proportional to the field amplitude and inversely proportional to the square of the beam  waist, which in the notations of this work corresponds to the third estimate in \eqref{estimations a0=1}.  
  % Lagrange function of the charged particle in an external electromagnetic field \cite{dmitriev2022effect}. 
  This is a consequence of the weak dependence of the waves in the paraxial approximation on $z$, which in the limiting case of the plane waves leads to the conservation of a quantity $(p_z - \varepsilon / c)$, which is  becoming in this situation an integral of motion.
  According to $\eqref{estimations a0=1}$ in this regime
  \begin{equation} \label{estimate p_z}
    (p_z - \varepsilon / c) \Big|_{{t_\infty}} \sim p_z^{(2)} \Big|_{{t_\infty}}. %\ll \varepsilon^{(4)} / c \Big|_{{t_\infty}},
  \end{equation}
  %where $\tau \sim \tau_{int}$ and  $\omega_0 \tau \gg 1$ were taken into account.
  To make the analysis consistent, the energy scaling in \eqref{estimate p_z} should not be greater, than that for the momentum, % in \eqref{estimations a0=1}, 
  %However, $\eqref{estimate p_z}$ cannot be satisfied if $p_z \Big|_{{t_\infty}} \ll \varepsilon / c \Big|_{{t_\infty}}$ as long as $\varepsilon / c \Big|_{{t_\infty}} \sim \varepsilon^{(4)} / c \Big|_{{t_\infty}}$.
  which is an omen, that higher orders of perturbation theory should be considered for $p_z$ in this regime. Indeed, in the frameworks of the paraxial and slowly varying temporal approximation for the discussed parameters the leading term in the expression for the fourth order longitudinal momentum \eqref{q1} may be estimated as $\sim m_e c^3 \frac{1}{w_0^2} \tau_{int}^2$ and has the form
  \begin{equation}
      p_z^{(4)}\left( {t_\infty}\right) = - \frac{m_e}{2} \int \limits_{-\infty}^{\infty} \frac{\partial}{\partial z_0} \left(\mathbf v^{(2)} (t)\right)^2 dt, 
  \end{equation}
  which after substituting of $\eqref{E_expanded}$ becomes
  \begin{equation}
      p_z^{(4)}\left( {t_\infty}\right) = \varepsilon^{(4)}\left( {t_\infty}\right) / c,
  \end{equation}
  due to the presence of $z$ in the envelope function $g(t - z/c)$.
  This expression reminds the integral of motion $p_z - \varepsilon / c$ for a charged particle in a plane wave.

  As a result, the estimated values of the gained momentum, energy and angular momentum for $a_0 \sim 1$, take form
  \begin{equation}
    \begin{aligned}
      \mathbf p_{\perp, \text{theor}} & = \mathbf p_\perp^{(2)}, \\
      p_{z, \text{theor}} & = p_z^{(2)} + \varepsilon^{(4)} / c, \\
      L_{z, \text{theor}} & = L_z^{(2)}, \\
      \langle L_{z, \text{theor}} \rangle_{\varphi_0} & = \langle L_z^{(4)} \rangle_{\varphi_0}, \\
      \varepsilon_{\text{theor}} & = \varepsilon^{(4)}, \\
    \end{aligned}
  \end{equation}
  where all the values are taken after the interaction at ${t_\infty}$ and $_\perp$ stands for the transverse components of vectors.
  It is possible, that these expressions turn to zero at some laser field configurations.
  In these cases higher orders of the perturbation theory or electromagnetic field expansions are required.
  However, configurations with non-zero values of these expressions definitely exist and will be presented further.
  
  \section{Some examples for certain polarization cases}\label{examples}

  \subsection{Circular polarization, one LG mode}
  Consider a circularly polarized Laguerre-Gaussian beam, $\sigma = \pm 1$.
  Boundary condition $\eqref{boundary_condition}$ takes the form
  \begin{equation}
    \begin{cases}
      \begin{aligned}
        \widetilde{H}_x \Big|_{z = 0} & = E_0 g \left( t \right) e^{i \omega_0 t} \frac{u_{pl}}{\sqrt{2}} \Big|_{z = 0}, \\
        \widetilde{H}_y \Big|_{z = 0} & = -i \sigma \widetilde{H}_x \Big|_{z = 0}.
      \end{aligned}
    \end{cases}
  \end{equation}

  As long as $u_{pl}$ depends on $\varphi$ as $\sim e^{- i l \varphi}$, the boundary condition depends on $\varphi$ as $\widetilde{H}_y \Big|_{z = 0} = -i \sigma \widetilde{H}_x \Big|_{z = 0} \sim e^{- il \varphi}$.
%  According to Appendix A, 
These relations between the components of the electromagnetic field apply to the solution of Maxwell equations in the whole space, i. e. for the solution in the whole space $\widetilde{H}_y = -i \sigma  \widetilde{H}_x \sim e^{- il \varphi}$.
   This leads to the following dependence of the exact solution on $\varphi$ in cylindrical coordinates
  \begin{equation}
    \begin{cases}
      \begin{aligned}
        \widetilde{H}_r & \sim e^{-i (l + \sigma)\varphi} \\
        \widetilde{H}_\varphi & = -i \sigma \widetilde{H}_r \sim e^{-i (l + \sigma)\varphi}, \\
        \widetilde{H}_z & \sim e^{-i (l + \sigma)\varphi}. \\
      \end{aligned}
    \end{cases}
  \end{equation}
  %\# \textcolor{red}{Deviations from these relations may appear as effects, not caught by the perturbation theory.}

  Hence, the components of $\mathbfcal{E}^{\{0\}}$ of the electric field $\eqref{E_expanded}$ are $\mathcal E_r^{\{0\}} \sim \mathcal E_\varphi^{\{0\}} \sim \mathcal E_z^{\{0\}} \sim e^{-i(l + \sigma) \varphi}$ and $\frac{\partial \left| \mathbfcal{E}^{\{0\}} \right|^2}{\partial \varphi_0} = 0$.
  Substituting this into $\eqref{Lz_particular_4_order}$ one obtains $\langle L_z^{(4)} \left( {t_\infty}\right) \rangle_{\varphi_0} = 0$, which means that the angular momentum in average is not transferred to electrons in the considered orders of the perturbation theory and the paraxial and slowly varying temporal envelope approximations.
  
  \subsection{Linear polarization, one LG mode}

  Consider a linearly polarized Laguerre-Gaussian beam, $\sigma = 0$.
  The boundary condition $\eqref{boundary_condition}$ takes the form
  \begin{equation}
    \begin{cases}
      \begin{aligned}
        \widetilde{H}_x \Big|_{z = 0} & = E_0 g \left( t \right) e^{i \omega_0 t} u_{pl} \Big|_{z = 0}, \\
        \widetilde{H}_y \Big|_{z = 0} & = 0.
      \end{aligned}
    \end{cases}
  \end{equation}

  In the lowest order of the paraxial approximation $\frac{\partial \left| \mathbfcal{E}^{\{0, 0\}} \right|^2}{\partial \varphi_0} = 0$, and the angular momentum in average is not transferred to electrons in the considered orders of approximations.
  
  \subsection{Linear polarization, superposition of LG modes}
  Probably one of the most efficient angular momentum transfer occurs in the configuration of electromagnetic fields, represented by a superposition of Laguerre-Gaussian modes with different azimuthal indexes, it provides a non-zero average gain of the angular momentum already in the lowest orders of approximations considered in this work.
  % as long as this effect may be described in the frameworks of the fourth order perturbation theory in the lowest possible orders of paraxial and slowly varying envelope approximation orders without additional expansions of electromagnetic fields on $\frac{1}{\omega_0 \tau}$ and $\frac{\lambda_0}{w_0}$.
  
  % In the cases of the linear and circular polarizations, on the opposite, additional smallnes appears, and hence, higher numerical accuracy is required both in calculations of electromagnetic fields and in the process of averaging of the gained angular momentum. 
  % Latter means more particles are required in the simulation.
  % If the numerical accuracy is not high enough, simulations become noisy and realistic average angular momentum transfer can not be obtained.
  
  Consider the case of linear polarization $\sigma = 0$, and a superposition of Laguerre-Gaussian beams with azimuthal indexes $l \neq m$. The boundary conditions $\eqref{boundary_condition}$ then may be written as
  \begin{equation} \label{boundary_lin_comb}
    \begin{cases}
      \begin{aligned}
        \widetilde{H}_x \Big|_{z = 0} & = E_0 g \left( t \right) e^{i \omega_0 t} \frac{u_{pl} + u_{qm}}{\sqrt{2}} \Big|_{z = 0}, \\
        \widetilde{H}_y \Big|_{z = 0} & = 0,
      \end{aligned}
    \end{cases}
  \end{equation}
  and in the lowest orders of the used approximations
  \begin{equation}
      \mathbfcal{E}^{\{\mathbf{0}, 0\}} = \frac{u_{pl} + u_{qm}}{\sqrt{2}} \mathbf e_x.
  \end{equation}

  % Due to interference between Laguerre-Gaussian modes of different azimuthal indexes, angular momentum transfer may be described in the lowest orders of paraxial and slowly varying envelope approximations.

  A single particle gains an angular momentum
  \begin{equation}
      L_z^{(2)}({t_\infty}) = - m_e c^2 \frac{a_0^2}{4} (l - m) \text{Re}\left( i u_{pl}^* u_{qm} \right) \tau_{int}.
  \end{equation}
  From this expression extreme values of gained angular momentum follow to be
  \begin{equation}\label{Lextr}
      L_{z, extr}^{(2)}({t_\infty}) = \pm m_e c^2 \frac{a_0^2}{4} (l - m) \big| u_{pl} u_{qm} \big| \Big|_{\varphi_0 = 0} \tau_{int}.
  \end{equation}
  % Note that because the intensity of the laser beam depends on the azimuthal angle $\varphi$, the averaging over initial angles of the particles for momentum and kinetic energy gains in this case are required.
  The expression for the average gained angular momentum reads after integration over $\varphi_0$
\begin{widetext}
  \begin{equation} \label{Lz_lm}
    \langle L_z^{\left(4\right)} \left( {t_\infty}\right) \rangle_{\varphi_0} = - \frac{m_e c^4}{\omega_0} (l - m) \frac{a_0^4}{64} \text{Re} \left\{ i \left( \frac{\partial A}{\partial r_0} \frac{\partial B^*}{\partial r_0} + \frac{(l - m)^2}{r_0^2} A B^* + 2 A \frac{\partial A^*}{\partial z_0}\right)\right\} \tau_{int}^2,
  \end{equation}
  where $A = u_{qm}^* u_{pl} \Big|_{\varphi_0 = 0}$ and $B = \frac{i z_0}{2} \left( u_{qm}^* \frac{\partial u_{pl}}{\partial z_0} - u_{pl} \frac{\partial u_{qm}^*}{\partial z_0} \right) \Big|_{\varphi_0 = 0}$.
  The average gained energy and longitudinal momentum read
  \begin{equation}
    \langle \varepsilon^{\left(4\right)} \left( {t_\infty}\right) \rangle_{\varphi_0} = m_e c^4 \frac{a_0^4}{128} \left( \left( \frac{\partial |u_{pl}|^2}{\partial r_0} + \frac{\partial |u_{qm}|^2}{\partial r_0}\right)^2 + 2 \Big|\frac{\partial \left( u_{pl} u_{qm}^*\right)}{\partial r_0} \Big|^2\right)\Big|_{\varphi_0 = 0} \tau_{int}^2,~~~ l \neq m,
  \end{equation}
\end{widetext}
  \begin{equation}
    \langle p_z^{\left(2\right)} \left( {t_\infty}\right) \rangle_{\varphi_0} = - \frac{m_e c^3}{\omega_0} \frac{a_0^2}{8} \frac{\partial \left( |u_{pl}|^2 + |u_{qm}|^2\right)}{\partial z_0} \Big|_{\varphi_0 = 0} \tau_{int}.
  \end{equation}

  Consider some special cases.
  \begin{itemize}
  \item One can see that when $l = m$ angular momentum is not transferred in the used orders of approximations.
  \item  If $p = q, \ m = -l$, both $A$ and $B$ become real and the average angular momentum turns to zero.
  It is a natural result, as far as such an electromagnetic field configuration does not carry an angular momentum.
  \item  If one changes azimuthal indexes $(l, m)$ to $(-l,-m)$, one can see that both $A$ and $B$ do not change, and the average gained angular momentum changes its sign which is also expected, because the OAM of the field also changes its sign.
  \end{itemize}
  
  % and next orders of perturbation theory or field expansion are needed to describe angular momentum transfer.
 
  To compare the obtained analytical estimates with numerical results, which may with a limited numerical accuracy provide both the electromagnetic fields and the angular momentum gained by particles beyond the perturbation theory limitations, the interaction of electrons with electromagnetic fields was studied with the PIC code Smilei \cite{Derouillat2018} in cylindrical geometry. To focus at the single-particle effects, the interaction between the electrons was not calculated. The use of the PIC code allowed to optimize the simulation time and, what is more important, to obtain the numerically calculated fields, which contains all the paraxial and temporal corrections.

  In simulations, the numerical box consisted of $6000$ cells in longitudinal direction and $1600$ in radial direction with spatial resolution of $2.5 \ nm$.  The laser field was injected from $z = 0$ boundary, with conditions $\eqref{boundary_lin_comb}$ placed at the boundary.
  The carrier laser frequency was $\omega_0 = 2.3 \times 10^{15} s^{-1}$, the beam waist radius was $w_0 = 1.3 \ \mu m$, the duration of the laser pulse $\tau = 2 \pi n / \omega_0$, where $n = 6$ is the number of periods.
  The dimensionless intensity used in the simulations was $a_0 = 1$, the radial indexes were $p = q = 0$ and three cases for azimuthal numbers were considered: $l = 0$, $m = 1$; $l = 1$, $m = 2$; and $l = 1$, $m = 3$.
  The common temporal envelope is chosen to be $g \left( t\right) = \cos^2\left(\frac{t - \tau/2}{\tau} \pi\right)$ when $|t - \tau/2| < \tau / 2$ and $0$ otherwise.
  In this case $\tau_{int} = \frac{3}{8} \tau$.
  Absorbing boundary conditions were set for the electromagnetic fields.
  For better statistics, electrons were initialized at $2.5 \ \mu m$ from the laser injection plane at given radial distances, but with different randomly distributed angles. These distances were defined as a set of fixed ten values from $r_0=0$ to $r_0=1.9 \ \mu m$ inside a thin disk with the axis being that of the laser beam.

\begin{figure}
    \centering
    \includegraphics[width = \linewidth]{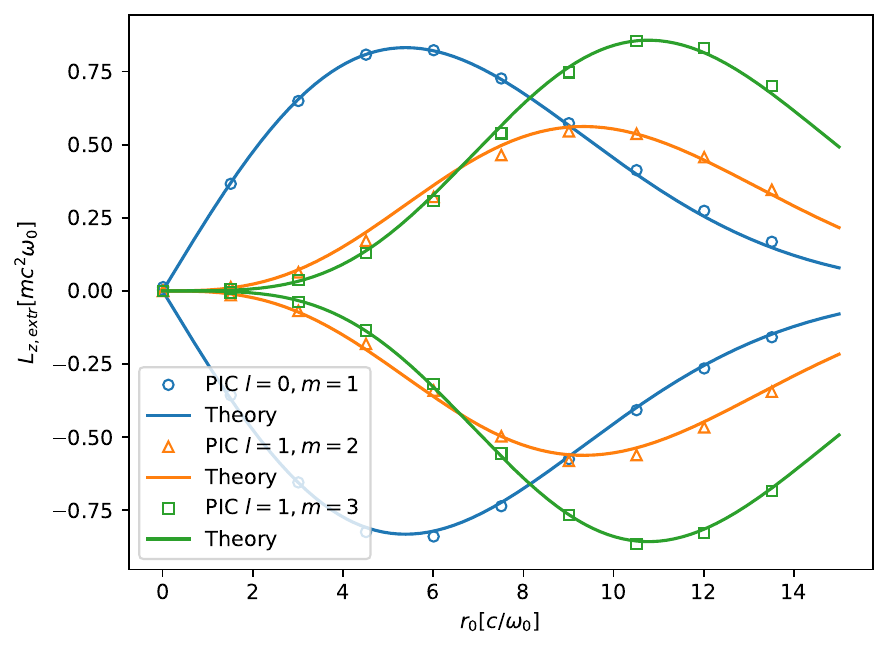}
    \caption{\footnotesize Extreme values of gained angular momentum after the interaction with (blue) $l = 0$, $m = 0$, (orange) $l = 1$, $m = 2$, (green) $l = 1$, $m = 3$ beams as a function of initial distance of 
      the particle from the beam axis.
      Dots, triangles and squares represent PIC numerical results, lines represent predictions of the model for the corresponding beam configurations.
    }
    \label{extr_Lz}
  \end{figure}

  Figure \ref{extr_Lz} represents the extreme values of the gained angular momentum after the interaction with three different beam configurations, the points are numerical results, the lines are drawn according to \eqref{Lextr}.
  Vertical axis corresponds to maximum and minimum extreme values of angular momentum, gained by particles, while horizontal axis corresponds to the initial radial distance of the particles from the beam axis.
  
  As long as the maximum local intensity of a Laguerre-Gaussian beam increases with the growth of the azimuthal index and its position shifts to larger radial distances, the maximum of the extreme value of the gained angular momentum also shifts toward the higher values of the initial distance from the beam axis with the growth of the azimuthal index.

  The average angular momentum transfer is an effect of the forth order perturbation theory while the extreme values of the gained angular momentum, indicating single particles gain, appears to be nonzero already in the second order.
  This leads to substantial statistical errors in calculations of the average angular momentum gain even when using a large amount of particles in the simulations.
  The average angular momentum, gained by the particles in the numerical simulations was calculated according to the following procedure
  \begin{equation}
      \overline{L_z}(r_0) \equiv \frac{\sum \limits_{i = 1}^{N_{r_0}} L_{z, i}}{N_{r_0}},
  \end{equation}
  where $L_{z, i}$ is the gained angular momentum of the i-th particle, which initial distance from the beam axis in the interval $[r_0, r_0 + \Delta r_0]$ and $N_{r_0}$ is the number of particles with initial coordinates in this interval.
  The error was estimated according to the central limit theorem
  \begin{equation} \label{Lz error}
      \Big| \overline{L_z} - \langle L_z \rangle_{\varphi_0} \Big| \sim \frac{\sqrt{\langle \left( L_z - \langle L_z \rangle_{\varphi_0} \right)^2 \rangle_{\varphi_0}}}{\sqrt{N}},
  \end{equation}
  where $\overline{L_z}$ stands for averaging over the ensemble of the particles in the numerical simulations and $\langle \dots \rangle_{\varphi_0}$ for the averaging over analytical expressions.
  
  As long as $\langle \left( L_z - \langle L_z \rangle_{\varphi_0} \right)^2 \rangle_{\varphi_0} = \langle L_z^2 \rangle_{\varphi_0} - \langle L_z \rangle_{\varphi_0}^2 < \langle L_z^2 \rangle_{\varphi_0}$ and
  $\langle L_z \rangle_{\varphi_0}$ is of the forth order and therefore may be considered negligible here compared to $\langle L_z^2 \rangle_{\varphi_0}^{1/2}$, the right hand-side may be estimated as
  \begin{multline}
      \sqrt{\langle \left( L_z - \langle L_z \rangle_{\varphi_0} \right)^2 \rangle_{\varphi_0} / N} \sim \sqrt{\langle \left(L_z^{(2)}\right)^{2} \rangle_{\varphi_0} / N} \sim \\
      \Big| L_{z, extr}^{(2)}({t_\infty}) \Big| / \sqrt{N} \sim m_e c^2 a_0^2 \tau_{int} / \sqrt{N}.
  \end{multline}

\begin{figure}
    \centering 
    \includegraphics[width=\linewidth]{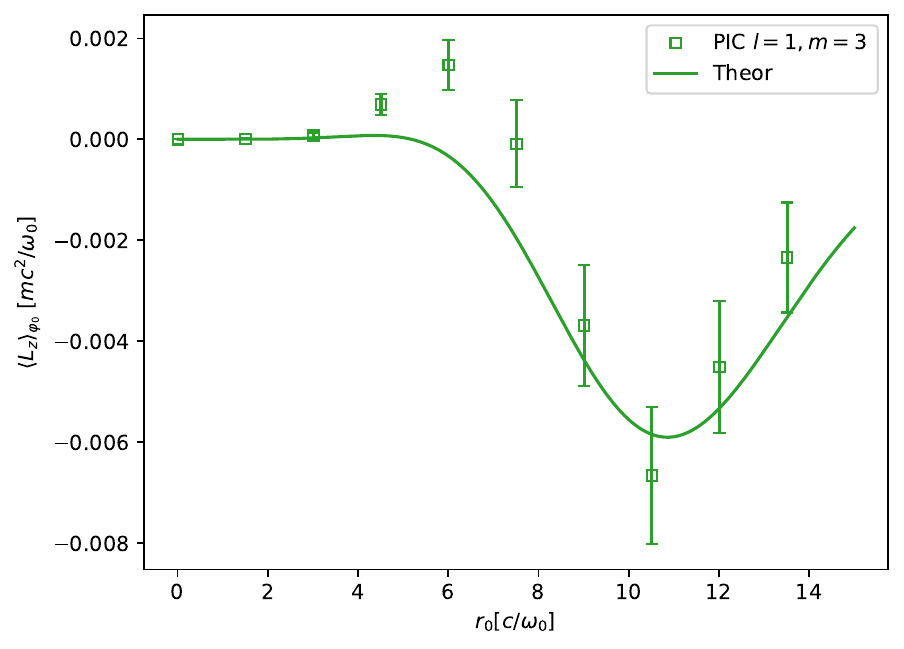}
    \caption{\footnotesize Average angular momentum, gained by electrons after the interaction with $l = 1$, $m = 3$ beams as a function of initial distance of 
      the particle from the beam axis.
      Squares represent PIC numerical results, solid line represents predictions of the model, vertical lines represent value of the statistical error.}
    \label{fig:Lz error}
  \end{figure}

\begin{figure}
    \centering 
    \includegraphics[width=\linewidth]{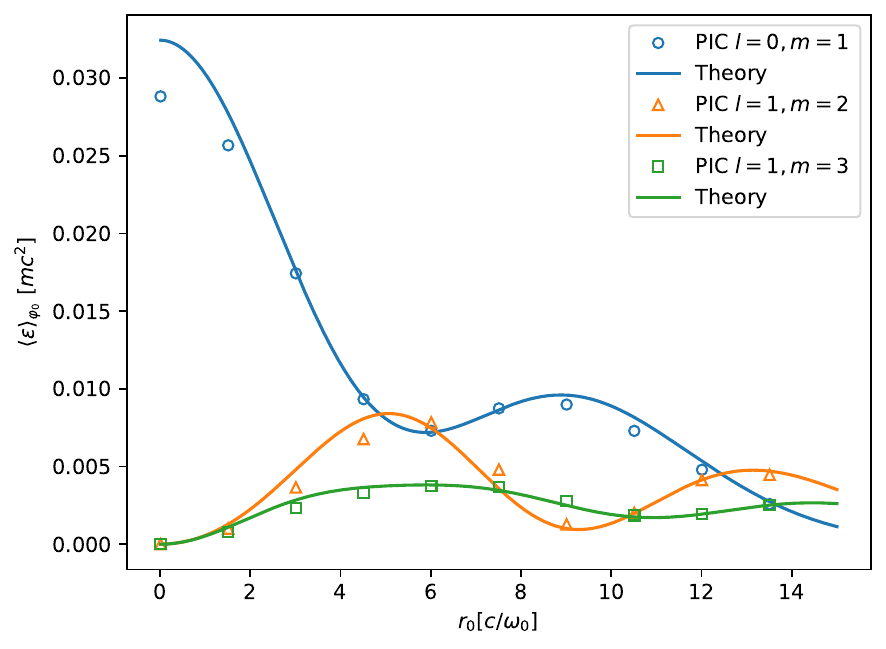}
    \caption{\footnotesize Average kinetic energy, gained by electrons after the interaction with (blue) $l = 0$, $m = 0$, (orange) $l = 1$, $m = 2$, (green) $l = 1$, $m = 3$ beams as a function of initial distance of 
      the particle from the beam axis.
      Dots, triangles and squares represent PIC numerical results, lines represent predictions of the model for the corresponding beam configurations.}
    \label{fig:ekin}
  \end{figure}

\begin{figure}
    \centering 
    \includegraphics[width=\linewidth]{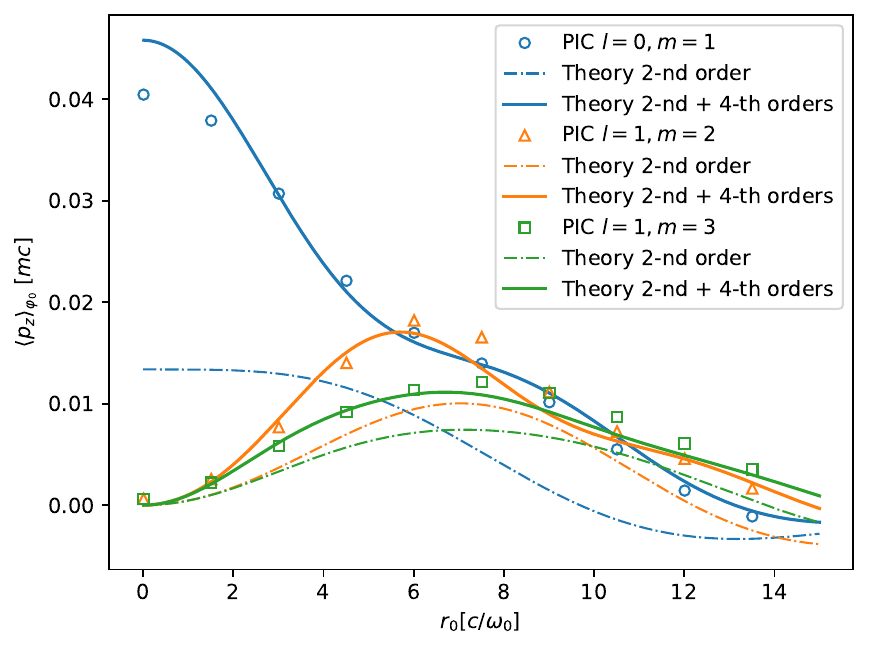}
    \caption{\footnotesize Average longitudinal momentum, gained by electrons after the interaction with (blue) $l = 0$, $m = 0$, (orange) $l = 1$, $m = 2$, (green) $l = 1$, $m = 3$ beams as a function of initial distance of 
      the particle from the beam axis.
      Dots, triangles and squares represent PIC numerical results, solid lines represent predictions of the model for the corresponding beam configurations, dashed lines represent predictions of the model without fourth order perturbation theory.}
    \label{fig:pz}
  \end{figure}

\begin{figure}
    \centering 
    \includegraphics[width=\linewidth]{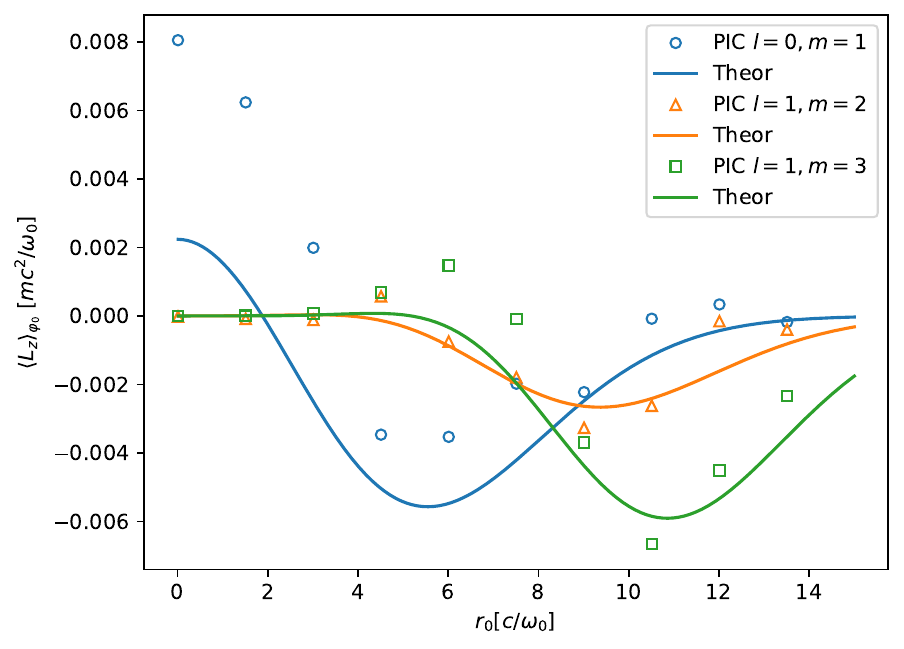}
    \caption{\footnotesize Average angular momentum, gained by electrons after the interaction with (blue) $l = 0$, $m = 0$, (orange) $l = 1$, $m = 2$, (green) $l = 1$, $m = 3$ beams as a function of initial distance of 
      the particle from the beam axis.
      Dots, triangles and squares represent PIC numerical results, lines represent predictions of the model for the corresponding beam configurations.}
    \label{fig:Lz}
  \end{figure}

  Relative statistical error is then may be estimated as
  \begin{equation}
      \frac{\sqrt{\langle \left( L_z - \langle L_z \rangle_{\varphi_0} \right)^2 \rangle_{\varphi_0} / N}}{\Big| \langle L_z^{(4)} \rangle_{\varphi_0}\Big|} \sim 
      \frac{\omega_0}{c^2}\frac{ w_0^2}{a_0^2 \tau_{int} \sqrt{N}},
  \end{equation}
  which rises when amplitude $a_0$ decreases. In the simulation results shown below, these errors are not always shown to make the plots readable, though their values are sometimes considerable. The exemplary plot in Fig. \ref{fig:Lz error} demonstrates the scale of the error bars for the case $l=1,m=3$. 
  Amount of the averaged over the initial azimuthal angle quantities gained by electrons, such as the energy, the longitudinal and the angular momentum versus their initial radial distance from the axis of the laser beam are shown in Figs. \ref{fig:ekin}, \ref{fig:pz}, \ref{fig:Lz}.
  % Statistical error is plotted in Fig. \ref{fig:Lz error} for gained angular momentum according to $\eqref{Lz error}$ for $l = 1$, $m = 3$ beam.

  The presented plots demonstrate that the perturbation theory allows a semi-quantitative description for the angular momentum gain by particles even at relatively high intensities, i.e. when $a_0 \sim 1$. As explained in the previous section, the reason for this is probably that the beam waist is large enough, i.e. $w_0/\lambda_0\gg1$.
  % The inverse of this parameter is related to the inhomogenousness of the laser beam and hence appears as an additional smallness in the perturbation theory, which may extend the limits of application of the model.

  \section{Discussion}\label{discussion}

  As the presented analysis of the angular momentum transfer shows, three parameters of the interaction are essential: dimensionless field amplitude $a_0$, relation of the period of laser oscillations to the laser pulse duration  $(\omega_0 \tau)^{-1}$ and the relation of the laser wavelength to the laser beam waist radius $\lambda_0 / w_0$.
  In this work, all these three parameters are considered as being small, i.e. $a_0 \ll 1$, $(\omega_0 \tau)^{-1} \ll 1$ and $\lambda_0 / w_0 \ll 1$, which allows to develop the perturbation theory based on expansion of the calculated values in powers of these small parameters.
  The relation between the parameters are not discussed, as the solutions are obtained in the lowest orders, which may provide a non-zero result. However, in general, especially, when the considered approximation gives a zero gained orbital momentum, e.g. the considered case of a single linearly polarized Laguerre-Gaussian beam, may require further expansion in $(\omega_0 \tau)^{-1}$ or $\lambda_0 / w_0$, depending on the relation between these parameters.

  It is interesting to note once again, that, according to the estimates \eqref{estimations}, the obtained expansion actually develops on the combination of these parameters, which at the same time takes into account the amplitude of the field, the beam waist, and the temporal behaviour. The combination is the relation of the particle oscillation amplitude squared to the beam waist radius squared and multiplied be the duration of the beam. This allows to consider as a result the near-relativistic values of $a_0\sim1$ and obtain a good semi-quantitative agreement between the analytical and the numerical results. It is also important, that the approximations used are not suitable for short laser pulses, where e.g. the phase effects may become important, as these effects break the symmetry over the azimuthal angle and facilitate the average OAM transfer.

  Discussions concerning the absorbed angular momentum in literature, usually start with defining a laser wave with a good-defined OAM, which are Laguerre-Gaussian beams. It is important however to take into account the corrections to the lowest orders of the paraxial and the slowly varying envelope approximations.
  Consider the simulations parameters, discussed in the previous section, and use the approximations for the electromagnetic field, rather than the numerical solution of the Maxwell with the algorithms provided by the PIC code.
  Namely, take the first orders of the paraxial and the slowly varying envelope approximations. The equations of motion were then solved numerically in a developed python code for individual particles, distributed as in previous simulations.
  In the first simulations consider the electromagnetic field as
  \begin{equation}
    \begin{cases}
      \begin{aligned}
        H_x & = - E_y & = & E_0 g(t - z) \text{Re}\left(e^{i\omega_0(t - z/c)} \frac{u_{pl} + u_{qm}}{\sqrt{2}}\right) \\
        H_y & = E_x & = & 0,
      \end{aligned}
    \end{cases}
  \end{equation}
  which is the main paraxial approximation and the main envelope approximation for the boundary conditions $\eqref{boundary_lin_comb}$.
  \begin{figure}[H]
    \centering 
    \begin{subfigure}[t]{0.45\textwidth}
        \centering
        \includegraphics[width = 1.01\textwidth]{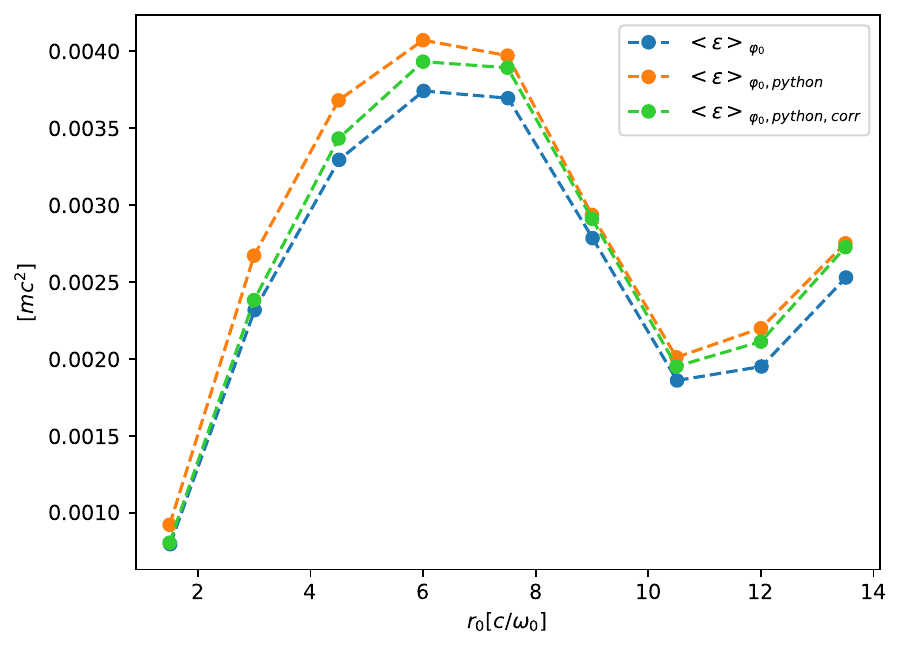}
    \end{subfigure}
    \hfill
    \begin{subfigure}[t]{0.45\textwidth}
        \centering
        \includegraphics[width = \textwidth]{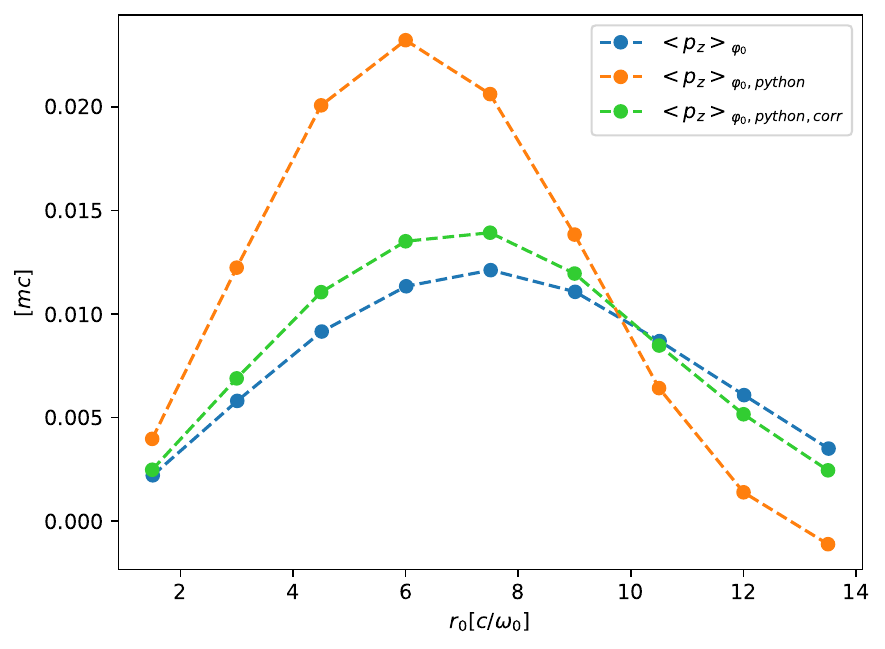}
    \end{subfigure}
    \hfill
    \begin{subfigure}[t]{0.45\textwidth}
        \centering
        \includegraphics[width = \textwidth]{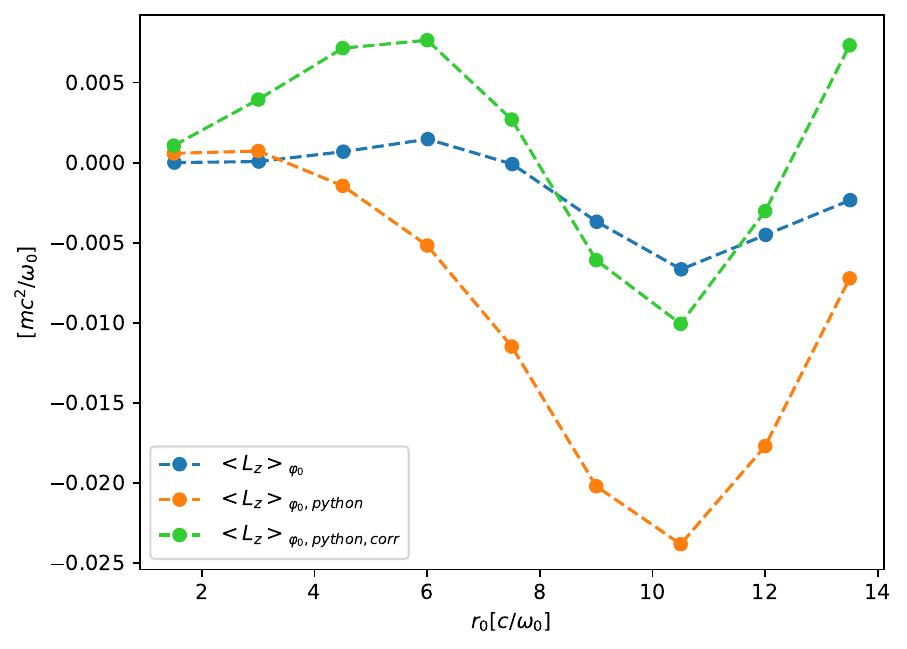}
    \end{subfigure}
    \caption{\footnotesize Average (a) kinetic energy, (b) longitudinal and (c) angular momenta gained by electrons after the interaction with $l = 1$, $m = 3$ beam as a function of initial distance of 
      the particle from the beam axis.
      Blue dots represent PIC code numerical results, orange lines represent python code numerical results with the lowest orders approximations.
      Green lines represent python code numerical results with the first corrections to the lowest orders approximations.}
    \label{comb_13_python}
  \end{figure}
  The other considered simulation takes into account the corrections according to 
  \begin{equation}\label{corrections_em}
    \begin{cases}
      \begin{aligned}
        \mathbf H & = E_0 \text{Re} \left\{\left( g(t - z/c) \left(\mathbfcal{H}^{\{\mathbf 0, 0\}} + \mathbfcal{H}^{\{\mathbf 0, 1\}} + \right. \right. \right.\\
        & \left. \left. \left. \mathbfcal{H}^{\{\mathbf 0, 2\}} + \mathbfcal{H}^{\{\mathbf 0, 3\}}\right) \right. \right. \\
        & \left. \left. + \frac{1}{\omega_0}g^\prime(t - z/c) \left( \mathbfcal{H}^{\{\mathbf 1, 0\}} + \mathbfcal{H}^{\{\mathbf 1, 1\}} \right) \right) e^{i\omega_0(t - z/c)}\right\} \\
        \mathbf E & = E_0 \text{Re} \left\{\left( g(t - z/c) \left(\mathbfcal{E}^{\{\mathbf 0, 0\}} + \mathbfcal{E}^{\{\mathbf 0, 1\}} + \right.\right.\right.\\
        & \left.\left.\left.\mathbfcal{E}^{\{\mathbf 0, 2\}} + \mathbfcal{E}^{\{\mathbf 0, 3\}} \right) \right. \right. \\
        & \left. \left. + \frac{1}{\omega_0}g^\prime(t - z/c) \left( \mathbfcal{E}^{\{\mathbf 1, 0\}} + \mathbfcal{E}^{\{\mathbf 1, 1\}} \right) \right) e^{i\omega_0(t - z/c)}\right\},
      \end{aligned}
    \end{cases}
  \end{equation}
  where the first three terms of expansion in powers of $\lambda_0/w_0$ and the first two terms of expansion in powers of $(\omega_0\tau)^{-1}$ are taken into consideration.

  The obtained after numerical integration gained average values, i.e. the kinetic energy, the longitudinal and the angular momentum of electrons are shown in Figs. \ref{comb_13_python}.
  One can see, that the longitudinal momentum and the angular momentum are rather different from the results, obtained with use of the PIC-calculated fields, if only the leading (lowest) orders of the approximations for the wave is used. Instead, the consideration of the wave with the corrections \eqref{corrections_em} results in a much better agreement between the values obtained with the use of the PIC code. Moreover, it is quite interesting to note that the gained momentum and angular momentum considerably decrease when the corrections are taking into account. So, the use of a rough approximation for the electromagnetic fields may result in a substantial overestimation of the average gained quantities.  

  %In the framework of the slowly varying envelope approximation odd orders of the perturbation theory result in zero average momentum, energy and angular momentum gain, as long as they contain products of odd number of oscillating functions, which average to zero during multiple oscillations of electromagnetic field.

  % \# in which case? 
  It is worth noting that within the used approach for the angular momentum gain \eqref{Lz_particular_4_order}, it appears that the first order correction of the slowly varying envelope approximation may has the contribution of the order $m_e c^4 a_0^4 \tau_{int}^2 / w_0^2 \omega_0$, which is the same as the contribution of the previous orders. 
  This is due to the fact that the lowest order of this approximation turns to zero after averaging over azimuthal angle $\varphi_0$.

  Interesting to note, that the condition \eqref{condition r1}, used to obtain \eqref{<Lz4>}, is satisfied for the lowest orders of paraxial and slowly varying approximations of the electromagnetic fields arising from the boundary condition $\eqref{boundary_condition}$ with $g \left( t\right) = \cos^2\left(\frac{t - \tau/2}{\tau} \pi\right)$ when $|t - \tau/2| < \tau / 2$ and $0$ otherwise.
  When the envelope function is $g(t) = \frac{1}{\sqrt{2 \pi}}e^{- (t - t_0)^2 / 2 \tau^2}$, this condition for the lowest orders of paraxial and slowly varying approximations is satisfied with exponential accuracy.

  However, the condition $\eqref{condition r1}$ is not always satisfied, e.g. for unipolar plane waves, such as $\mathbf E = E_0 e^{- (t - z)^2 / 2 \tau^2} \mathbf e_x$, $\mathbf H = E_0 e^{- (t - z)^2 / 2 \tau^2} \mathbf e_y$.
  Such plane waves %are only an approximation to a more realistic electromagnetic field configuration, and hence 
  are not discussed in this paper.

  % This results in the scaling of gained energy, momentum and angular momentum as $a_0^2$ and not as $a_0$, and the next non-vanishing orders of perturbation theory add $a_0^2$ to the previous value.
  % It is also notable, that any new order of perturbation theory adds $\lambda_0/w_0$ as a result of inhomogenousness of electromagnetic fields, and hence, gained energy, momentum and angular momentum get additional $\lambda_0^2 / w_0^2$ every non-vanishing order perturbation theory.
  % Integration over time also adds $\tau_{int}$ for every even order of perturbation theory, while every odd vanishes, as was stated earlier.
  % Hence it is natural to expect that perturbation theory for gained energy, momentum and angular momentum has another smallness parameter, different from $a_0$, which is probably $\lambda_0 c \frac{a_0^2 \tau_{int}}{w_0^2}$.
  % As a result, it is possible that the expressions for gained energy, momentum and angular momentum may still be applicable at amplitudes $a_0 \sim 1$ due to additional smallness of $\lambda_0 c \frac{\tau_{int}}{w_0^2}$.
  % An exception arises in longitudinal momentum $p_z$, which also requires consideration of the fourth order term $\varepsilon^{(4)}/c$, which will be discussed further.

  Intensity of a Laugerre-Gaussian beam has a form of several rings with a center on the axis of the beam and radius $\sim w_0$.
  This allows one to assume that electrons gain angular momentum mostly in the regions of these rings and hence, form a solenoid with a characteristic radius $\sim w_0$.
  According to the model of generation of the magnetic field by a solenoid with charged current \cite{nuter2018plasma}, the magnetic field value may be estimated (in CGS units) as $H \sim e n_0 w_0 v_\varphi / c \sim \frac{e n_0 c^3 a_0^4 \tau_{int}^2}{\omega_0 w_0^2}$.
  For the parameters, normal for modern laser experiments, $\omega_0 = 2.3 \times 10^{15} s^{-1}$, $w_0 = 1.3 \ \mu m$, $\tau = 12 \pi / \omega_0$, $a_0 = 1$, $n_0 = 0.01 \ n_c$, where $n_c$ is the critical plasma density, magnetic field may be estimated as $H \sim 60 \ T$.
  % \textcolor{red}{In dimensionless units $H \sim n_0 a_0^3 \tau^2 / w_0^2$, where $n_0$ is measured in critical densities. In our parameters $H \sim 0.01$}

It should be noted, that electrons mostly gain angular momentum of the opposite sign to that of the laser beam. The direction of the rotation of the electrons corresponds to the generated magnetic field directed along the laser beam propagation direction. This is an interesting results, which also was observed in a full scale 3D PIC simulations \cite{nuter2018plasma, nuter2020gain}.

In this work, no collective effects were considered. Of course, the collective effects, as well as effects which dephase the particle motion in the wave, such as collisions, ionization, radiation friction and others, may qualitatively change the interaction process, though it is easy to find the conditions when the single-particle processes dominate. What is actually done in the work is the initial step towards the understanding of the OAM transfer from light waves to single particles, in situation when the particle distribution is isotropic so that there is no initial axial asymmetry in the system except the laser wave phase. The obtained important result is that indeed, in this situation the OAM may be transferred to the particles, the process efficiency is growing with the increasing of the field amplitude and decreasing the beam waist.

  % \textcolor{red}{Add description and discussion.}
  
  %In some special cases additional terms of the expansion may be required to correctly describe the process of the angular momentum transfer, especially in the cases, briefly discussed in this paper, such as linear polarization, where additional expansions in powers of $\lambda_0 / w_0$ and $1/\omega_0 \tau$ may be required and the amount of the terms required is determined by the relation between these parameters.

    %Main idea:
    %Paraxial approximation and SWEA without corrections are not %enough both numerically and analytically in some occasions

    %How to illustrate?
    
    %Papers to discuss: 
    %1) \cite{nuter2020gain} -- calculations are beyond 
    %  application limits (fields are zero order 1/tau, however %1/tau terms are considered in calculations of Lx)
    %2) \cite{nuter2018plama} -- OAM gain as the 2-nd order, %fields 
    %  are phenomenological
    %3) 
  
  % \textcolor{red}{Discussion of errors of $L_z$ calculations?}

  % \textcolor{red}{
  %   Discuss necessity of accurate em field calculations.
  %   Make reference to other works.
  %   \begin{enumerate}
  %       \item Numerical: if one does not consider corrections, AM transfer may occur as the 2-nd order effect.
  %       Arises if one solve equations of motion for free electrons with pre-defined EM feilds.
  %       \item Analytical: AM transfer may occur as the 2-nd order, if one goes beyond the limits of applicability.
  %   \end{enumerate}
  % }
  
  \section*{Acknowledgements}
  
  The work was supported by the Foundation for the Advancement of Theoretical Physics and Mathematics “BASIS”.
  
  The calculations were performed on the hybrid supercomputer K60 installed in the Supercomputer Centre of Collective Usage of KIAM RAS.

%\bibliography{refs}% Produces the bibliography via BibTeX.

\pagebreak

\begin{widetext}

\appendix\label{Appendix}

  \section{Electromagnetic fields in paraxial and slowly varying envelope approximation} \label{EM_calculation}

      \subsection{Table integrals}

In calculation of the electromagnetic fields using the boundary condition at $z=0$, the integration over angles $\varphi$ results in the Bessel function 
\begin{equation}
    2\pi i^s J_s(z)=2\pi i^{|s|} J_{|s|}(z)=\int_0^{2\pi} d\phi \exp [i s \phi + i z \cos \phi]
\label{Bessel_representation}
\end{equation}
with integer $s$, 

\begin{equation}
    H_\alpha(\omega, \mathbf k_\perp, z=0) = 2 \pi w_0^2 g_{\omega - \omega_0} E_0 \sum \limits_{p, l} a_{pl\alpha} i^{|l|} e^{- i l \theta} \int \mathtt{r} d\mathtt{r} U_{pl} (\mathtt{r}, 0)  J_{|l|}( k_\perp r_\perp).
    \label{Bessel_in_Ealpha}
\end{equation}
% where $g_{\omega} = \int d\xi g\left( \xi \right) e^{-i \omega \xi}$ is the Fourier transform of the envelope and $\theta$ is the angle between $\mathbf k_\perp$ and $x$-axis, such that $\mathbf k_\perp \cdot \mathbf r_\perp = |\mathbf k_\perp| r \cos(\theta - \varphi)$.

%\bb{In $U_{pl}$ there is no $|l|$ in $e^{il\varphi}$ -- is it correctly counted in \eqref{E_alpha-from-boundary}}

Using the following identity from \cite{gradshteyntable} for an integral of Laguerre polynomials $ L_n^{s}(z)$ with Bessel functions $J_s(z)$ (here $n$, $s$, $a$, $b$ are parameters)
  \begin{equation}
      \int \limits_0^{\infty} x^{s + 1} e^{- b x^2} L_n^{s} \left( a x^2 \right) J_s (xy) dx = \frac{(b - a)^n}{2^{s + 1} b^{s + n + 1}} y^s e^{-\displaystyle\frac{y^2}{4b}} L_n^s \left( \frac{a y^2}{4 b (a - b)}\right)
      \label{ryzhik_identity}
  \end{equation}
  after integrating \eqref{Bessel_in_Ealpha} over $\mathtt{r}$ one obtains \eqref{E_alpha-from-boundary}.

      \subsection{Expressions for the fields in terms of the used perturbation theory} \label{Field_components}

  \begin{equation}
    \begin{aligned}
      \widetilde{\mathbf E}_{\perp}^{\{\mathbf 0, 0\}} & = E_0 g \left( \xi \right) \mathbfcal{E}_{\perp}^{\{\mathbf 0, 0\}} e^{i \omega_0 \xi}, \\
      \widetilde{\mathbf E}_{\perp}^{\{\mathbf 1, 0\}} & = \frac{E_0}{\omega_0} g^\prime \left( \xi \right) i z \frac{\partial \mathbfcal{E}_{\perp}^{\{\mathbf 0, 0\}}}{\partial z} e^{i \omega_0 \xi}, \\
      \widetilde{\mathbf E}_{\perp}^{\{\mathbf 0, 2\}} & =  \frac{E_0 c}{\omega_0} g \left( \xi \right) \left( \frac{z}{2i} \frac{\partial^2 \mathbfcal{E}_{\perp}^{\{\mathbf 0, 0\}}}{\partial z^2} - i \mathbf e_x \left( \frac{\partial \mathcal H_{z}^{\{ \mathbf 0, 1\}}}{\partial y} - \frac{\partial \mathcal H_y^{\{\mathbf 0, 0\}}}{\partial z}\right) - i \mathbf e_y \left( \frac{\partial \mathcal H_x^{\{\mathbf 0, 0 \}}}{\partial z} - \frac{\partial \mathcal H_{z}^{\{\mathbf 0, 1\}}}{\partial x}\right)\right) e^{i \omega_0 \xi}, \\
    \end{aligned}
  \end{equation}
  \begin{equation}
    \begin{aligned}
      \widetilde{H}_{z}^{\{\mathbf  0, 1 \}} & = E_0 g \left( \xi \right) \mathcal H_{z}^{\{\mathbf 0, 1\}} e^{i \omega_0 \xi}, \\
      \widetilde{H}_{z}^{\{\mathbf  1, 1 \}} & = \frac{E_0}{\omega_0} g^\prime \left( \xi \right) i \frac{\partial}{\partial z} \left( z \mathcal H_{z}^{\{\mathbf 0, 1\}} \right) e^{i \omega_0 \xi}, \\
      \widetilde{H}_{z}^{\{ 0, 3 \}} & = \frac{E_0 c}{\omega_0 }g \left( \xi \right) \left( \frac{z}{2i} \frac{\partial^2 \mathcal H_{z}^{\{\mathbf 0, 1\}}}{\partial z^2}  - i \frac{\partial \mathcal H_{z}^{\{\mathbf  0, 1\}}}{\partial z}  \right)e^{i \omega_0 \xi}, \\
    \end{aligned}
    \hspace{18pt}
    \begin{aligned}
      \widetilde{E}_{z}^{\{\mathbf 0, 1 \}} & = E_0 g \left( \xi \right) \mathcal E_{z}^{\{\mathbf 0, 1\}} e^{i \omega_0 \xi}, \\
      \widetilde{E}_{z}^{\{\mathbf  1, 1 \}} & = \frac{E_0}{\omega_0} g^\prime \left( \xi \right) i \frac{\partial}{\partial z} \left( z \mathcal E_{z}^{\{\mathbf 0, 1\}} \right) e^{i \omega_0 \xi}, \\
      \widetilde{E}_{z}^{\{\mathbf 0, 3 \}} & = \frac{E_0 c}{\omega_0} g \left( \xi \right) \frac{z}{2i} \frac{\partial^2 \mathcal E_{z}^{\{\mathbf  0, 1 \}}}{\partial z^2} e^{i \omega_0 \xi}, \\
    \end{aligned}
  \end{equation}
  where
  $\mathbfcal{E}_{\perp}^{\{\mathbf 0, 0\}} = -\mathbf e_z \times \mathbfcal{H}_{\perp}^{\{\mathbf 0, 0\}}$,
  $\mathcal H_{z}^{\{\mathbf 0, 1\}} = - i \frac{c}{\omega_0}\nabla_\perp \cdot \mathbfcal{H}_{\perp}^{\{\mathbf 0, 0\}}$,
  $\mathcal E_{z}^{\{\mathbf 0, 1\}} = -i \frac{c}{\omega_0} \left(\nabla_\perp \times \mathbfcal{H}_{\perp}^{\{\mathbf 0, 0\}} \right)_z$.

\end{widetext}

\newpage

\end{document}